\documentclass[universe,article,accept,pdftex,moreauthors]{Definitions/mdpi} 

\firstpage{1} 
\pubvolume{1}
\issuenum{1}
\articlenumber{0}
\pubyear{2023}
\copyrightyear{2023}
\externaleditor{{Academic Editor: Firstname Lastname} 
}
\datereceived{7 April 2023} 
\daterevised{6 June 2023} 
\dateaccepted{12 June 2023 } 
\datepublished{ } 
\hreflink{https://doi.org/} 

\usepackage{bm}
\usepackage{amsmath}
\usepackage{epsfig}
\usepackage{graphicx}
\usepackage{pdfpages}
\usepackage{color}   
\newcommand{\bfvec}[1]{\hbox{\boldmath$#1$\unboldmath}}

\newcommand{\QCD}{\mathrm{QCD}}
\newcommand{\eff}{\mathrm{eff}}

\Title{The Interrelated 
 Roles of Correlations in the Nuclear Equation of State and in Response Functions: Application to a Chiral Confining Theory}

\TitleCitation{The Interrelated Roles of Correlations in the Nuclear Equation of State and in Response Functions: Application to a Chiral Confining Theory}

\Author{Guy Chanfray 
  $^{1,}$*\orcidA{}, Magda Ericson $^{1,2}$ and Marco Martini $^{3,4}$\orcidC{} 
}

\AuthorNames{Guy Chanfray, Magda Ericson and Marco Martini}

\AuthorCitation{Chanfray, G.; Ericson, M.; Martini, M.}

\address{%
$^{1}$ \quad University Claude Bernard Lyon 1, CNRS/IN2P3, IP2I Lyon, UMR 5822, F-69622~
 Villeurbanne, France; mericson@cern.ch\\
$^{2}$ \quad Theory Division, CERN, CH-12111 Geneva, Switzerland 
 \\
$^{3}$ \quad IPSA-DRII,  63 boulevard de Brandebourg, 94200 Ivry-sur-Seine, France; marco.martini@ipsa.fr\\
$^{4}$ \quad Sorbonne Universit\'e, Universit\'e Paris Diderot, CNRS/IN2P3, Laboratoire
de Physique Nucl\'eaire et de Hautes Energies (LPNHE), Paris, France 
}

\corres{Correspondence: g.chanfray@ipnl.in2p3.fr}

\abstract{ We study the role of short-range correlations, as well as pion and rho loops governing long-range RPA correlations, in nuclear matter properties and  response functions. We use an adapted formulation of the Brueckner G-matrix approach to generate a pair correlation function satisfying the Beg--Agassi--Gal theorem, providing a natural cutoff to the loop integrals.  We present results for the case of a relativistic chiral theory, including the effects of quark confinement and of the chirally broken vacuum in a version where parameters are directly connected to QCD observables or constrained by well-established hadron phenomenology. This provides a unified and coherent view of the nuclear matter equation of state and the effect of correlations on neutrino--nucleus scattering. }


\begin{document}

\section{Historical Introduction}\label{Intro}
More than thirty years ago, the observed depletion of the nuclear structure function with respect to the free nucleon one, with the maximum effect observed when the Bjorken variable had a value of  $x\simeq 0.6$   \cite{EMC84,EMC88}, was tentatively attributed to a possible modification of the nucleon structure in the nuclear medium. The reason was the inability
of the standard mean-field Hartree--Fock approach \cite{Aku85} to reproduce this celebrated EMC effect, as discussed in \cite{LI88}. This depletion is governed by the separation energy, $\left\langle \epsilon\right\rangle$, a negative quantity that is  the opposite of the mean energy needed to remove a nucleon from the nucleus. For typical nuclei from  $^{12}$C to $^{56}$Fe, the value $\left\langle\, \epsilon\,\right\rangle=-(40 \div 50)$ MeV  gives a correct fit to the EMC effect. Hence, according to the Koltun sum rule, $-\left\langle\, \epsilon\,\right\rangle=\left\langle\, t\,\right\rangle +2\,B$ \cite{Ericson87}, this necessitates a nucleon mean kinetic energy as large as $\left\langle\, t\,\right\rangle \sim 35$ MeV, in strong disagreement with the conventional HF calculation giving $\left\langle\, t\,\right\rangle_{HF} \sim 20$ MeV. The reason for this large value of kinetic energy is well known and is linked to correlations. Realistic calculations based on correlated wave functions have shown that the bare nucleon momentum distribution acquires a long tail beyond the Fermi momentum since the effect of short-range correlations is the depopulation of the Fermi sea. For instance, in the evaluation by Ciofi degli Atti et al. \cite{Cioffi,Cioffi1,Cioffi2}, for nuclei ranging from $^{12}$C to $^{56}$Fe, the kinetic energy per nucleon becomes larger, namely, $\left\langle\, t\,\right\rangle\simeq 35$ MeV, and the absolute value of the separation energy also becomes  larger,  $-\left\langle\, \epsilon\,\right\rangle\simeq 50$ MeV. Thus, conventional nuclear effects are indeed able to reproduce the major part of the EMC effect.  One can  conclude that the independent particle picture is a very approximate view of the nucleus or that the Hartree--Fock scheme  with bare nucleons is a poor approximation of the nuclear ground state. This conclusion seems surprising at first since we know that independent particle models based on the HF scheme accurately reproduce many basic properties of nuclei. The reason for the apparent discrepancy is due  to the implicit identification of bare and dressed nucleons. In reality, Hartree--Fock nucleons should not be treated as bare nucleons but as nucleons surrounded by a polarization cloud, constituting  quasi-particles (in the Landau sense), moving independently in a smooth mean-field potential. As pointed out in \cite{Chanfray91}, such dressed objects couple differently from bare ones to an external (electroweak) probe.

A similar story occurred twenty  years later in 2009/2010, when the measurement of the charged-current quasielastic-like  (CCQE-like) neutrino--nucleus cross-section performed by the MiniBooNE collaboration \cite{Katori2009,MiniBooNe2010} turned out to be in strong disagreement with the predictions based on the Fermi gas model 
(the model implemented in all Monte Carlo at that time; see, e.g., \cite{MiniBooNe2010}), 
suggesting that scattering between a neutrino and independent nucleons strongly underestimates the measured cross-section. It was first proposed that this effect was due to a possible in-medium modification of the nucleon axial form factor, i.e., to an intrinsic modification of the bound nucleon structure. However, very soon,  the explanation of this disagreement was given by the  Lyon group \cite{MECM2009}: nucleons are correlated via short-range correlations (SRCs) and meson exchange currents (MECs), which implies the possible ejection of a pair of nucleons  from the scattering of the neutrino. This process is referred as the two-particle--two-hole (2p-2h) 
 or n-particle--n-hole (np-nh) contribution to the cross-section. As an elaboration of the approach outlined in \cite{Chanfray91}, we recently proposed in~\cite{CEM2022} a general technique based  on a unitary transformation mapping of the Fermion operators relative to bare nucleons into quasi-particle operators relative to dressed nucleons. In short, this unitary transform is governed by the G matrix, which is the effective interaction felt by the dressed HF nucleon,  with this G matrix $G(r)$ being very schematically related to the bare nucleon interaction $V(r)$ by $G(r)=V(r)f(r)$, where $f(r)$ is the pair correlation~function.

In the following, we elucidate the influence of short-range correlations, which depopulate the Fermi sea and enhance the response functions, on the nuclear matter equation of state. The discussion is specifically illustrated on the basis of a microscopic model, inspired by  QCD, describing the nucleon--nucleon interaction and its modification in the nuclear medium. 

In the second section, we  slightly reformulate the approach of \cite{CEM2022}, which allows the generation of the G matrix governing the interaction between dressed nucleons. We then propose a way to cause the resulting correlation function to vanish at zero distance, in agreement with the Beg--Agassi--Gal theorem \cite{Beg, Chanfray1985}. The parameter $q_C$ appearing in this correlation function represents the inverse of the correlation hole radius induced by short-range correlations (SRCs). Its value can be fixed  from the bare $NN$ interaction derived from  a relativistic chiral  theory, including the effect of  quark confinement, developed in the recent years  \cite{Chanfray2001,Chanfray2005,Martini2006,Chanfray2007,Massot2008,Massot2009,Chanfray2011,Massot2012,Rahul}.

This approach, which we now call the chiral confining model, is the subject of Section~\ref{sec3}. The original phenomenological version \cite{Chanfray2005,Chanfray2007,Massot2008} described in Section 
 \ref{sec3.1} is based on a linear sigma model  (L$\sigma$M) Lagrangian  with standard Yukawa couplings of the various meson fields (scalar $s$, $\pi,\,\rho,\,\omega, \ldots$) to 
 the nucleon but with three important physical pillars. First, the scalar field $s$, associated with the radial fluctuation of the chiral condensate, is identified with the sigma meson of relativistic theories \cite{Chanfray2001,Martini2006}. Second, the effect of the quark substructure  of the nucleon is reflected by its polarizability in the presence of the nuclear scalar field generating a repulsive three-nucleon force, providing an efficient saturation mechanism \cite{Chanfray2005,Chanfray2007,Massot2008,Rahul}. Third, the associated response parameters, namely, the nucleon scalar coupling constant, $g_S$, and scalar susceptibility, $\kappa_{NS}$, can be related to two chiral properties of the nucleon given by lattice QCD simulations, imposing severe constraints on their values \cite{Chanfray2007,Massot2008,Rahul}. This model has been  applied in the past to the equation of state of nuclear matter and neutron stars, as well as to the study of the chiral properties of nuclear matter at different levels of approximation in the treatment of the many-body problem (RMF, Relativistic Hartree--Fock, or RHF, pion loop correlation energy). 

However, it has been realized that the correct description of saturation properties compatible with lattice QCD constraints requires a large value of the scalar susceptibility, $\kappa_{NS}$, which is not compatible with realistic confining models of the nucleon. Hence,  in Section \ref{sec4.2}, we present an enriched version recently proposed in \cite{Chanfray2023}, the NJL chiral confining model, which is based on the semi-bosonization of the Nambu--Jona-Lasinio model, as described in \cite{Chanfray2011}. The main  new feature of this improved framework is the replacement of the $L\sigma M$ effective potential, $V_{\chi,L\sigma M}$,
by a more repulsive NJL one, $V_{\chi, NJL}$, 
 considerably reducing the tension between the expected model values of the response parameters and the lattice QCD constraints.

A further step  is accomplished in Section \ref{sec3.3}, where we build an effective Hamiltonian inspired by QCD that allows the simultaneous introduction, within some ansatz prescriptions, of a confining model for the nucleon, allowing the calculation of the response parameters, and an equivalent NJL model. This is the foundation of the QCD-connected  chiral confining model, which basically depends on two genuine QCD quantities, namely, the string tension, $\sigma$, and the gluon condensate, $\mathcal{G}_2$.

Section \ref{sec4} is devoted to nuclear matter calculations. We first summarize the values and the origin of the various parameters entering the (in-medium) nuclear interaction, trying to distinguish between those coming from the QCD-connected model (essentially the scalar and pionic sectors) and those coming from hadron phenomenology (essentially the vector sector), and we explain how the effect of short-range correlations is implemented in this context through the adapted G-matrix approach.  We  thus finally discuss the combined effect of short-range correlations and long-range RPA-like correlations (pion--rho loops) on nuclear matter properties and the depopulation of the Fermi sea, which is an indicator of the magnitude of the 2p-2h contribution to the neutrino--nucleus scattering cross-section. 

\section{Connecting Different Microscopic Models for Correlations}\label{sec2}
\subsection{From Bare to Dressed Nucleons: Landau Quasi-Particles }\label{sec2.1}
	Let us consider a nuclear system described by a Hamiltonian where the bare nucleon--nucleon interaction reduces to a two-body potential, written with standard notation as
 \begin{equation}
		H=\sum_{k_1, k_2}<k_1\,|\,T\,|\, k_2> a^\dagger_{k_1}a_{k_2} +\frac{1}{4} \sum_{k_1 k_2 k_3 k_4}<k_1\,k_2\,|\,\bar{V}\,|\,k_3\, k_4>
		a^\dagger_{k_1}a^\dagger_{k_2}a_{k_4}a_{k_3},
  \end{equation}
  where  $a$ ($a^\dagger$) is the annihilation (creation) operator relative to the bare nucleons. According to Goldstone \cite{Goldstone}, the correlated ground state of the nucleus, $|0>$, can be obtained from the state of uncorrelated bare nucleons, $|0>_{uncorr}$, by a unitary transformation as~follows:
	\begin{eqnarray}
		|0>&=&U\, |0>_{uncorr} = U(a) \,\left(\prod_h a_h^\dagger\right)\,|vac>=U(a) \,\left(\prod_h a_h^\dagger\right)\,U^\dagger(a)\,|vac>\nonumber\\
		&\equiv& \left(\prod_h A_h^\dagger\right)\,|vac>,
	\end{eqnarray}
where $|vac>$ is the true vacuum state with a baryonic number equal to zero. The label $h$ stands for hole states with a momentum below the Fermi momentum $k_F$.
Hence, the correlated ground state $|0>$ can be obtained as a Slater determinant of dressed nucleons created by the $A_h^\dagger$ operators. These operators are related to the bare creation operators, $a_h^\dagger$, by a unitary transformation preserving the anticommutation relation: 
	\begin{equation}
		A_h^\dagger=U(a)\, a_h^\dagger\, U^\dagger(a)\,\,\,\Leftrightarrow\,\,\,a_h^\dagger=U^\dagger(A)\, A_h^\dagger\, U(A).
	\end{equation}
	
 Starting, 
 as in Ref.~\cite{Desplanques}, with a unitary operator $U(A)=\exp(S(A))$, with $S(A)$ being an anticommutation operator truncated at 2p-2h excitations,
	\begin{equation}
		S(A)=\frac{1}{4}\left(s_{p_1 p_2 h_1 h_2} A^\dagger_{p_1}A^\dagger_{p_2}A_{h_1}A_{h_2}\,-\,hc\right),
	\end{equation}
	one obtains, to the leading order in $S(A)$, the following result  for hole ($h$) and particle ($p$) creation and annihilation operators:
	\begin{eqnarray}
		\label{eq_ah}
		a_h &=& \sqrt{Z_h}\,A_h \,-\,\frac{1}{2}\,\sum_{h_2 p_3 p_4}\,s_{p_3 p_4 h_2 h}\,A^\dagger_{h_2}\,A_{p_4} \,A_{p_3} \label{eq_ah}\\
		a_p &=& \sqrt{Z_p}\,A_p \,-\,\frac{1}{2}\,\sum_{p_2 h_3 h_4}\,s_{p p_2 h_3 h_4}\,A^\dagger_{p_2}\,A_{h_3} \,A_{h_4},
		\label{eq_ap}
	\end{eqnarray}
	where $Z_h$ and  $Z_p$ \textls[-15]{are normalization factors that are fixed to ensure that} $<0|\{a_k,a^\dagger_k\}|0>=1$:
	\begin{equation}
		Z_h = 1\,-\,\frac{1}{2}\,\sum_{h_2 p_3 p_4}\left|s_{p_3 p_4 h_2 h}\right|^2, \qquad
		Z_p = 1 \,-\,\frac{1}{2}\,\sum_{p_2 h_3 h_4}\left|s_{p p_2 h_3 h_4}\right|^2. 
		\label{Zfactor}
	\end{equation}
	
To the same order, the ground state energy can obtained  by a straightforward although tedious calculation as a functional of $s_{p_1 p_2 h_1 h_2}$. These coefficients can be obtained variationally, i.e., by minimizing the ground-state energy, yielding:
\vspace{-9pt}
\begin{adjustwidth}{-\extralength}{0cm}
\centering 
\begin{eqnarray}
 && \left(\epsilon_{h_1} +\epsilon_{h_2}-\epsilon_{p_1}-\epsilon_{p_2 }\right)s_{p_1 p_2 h_1 h_2} =<p_1\,p_2\,|\,\bar{V}\,|\,h_1\, h_2>\nonumber\\
 && + \frac{1}{4}\sum_{p'_1, p'_2} <p_1\,p_2\,|\,\bar{V}\,|\,p'_1\, p'_2>  s_{p'_1 p'_2 h_1 h_2}+\frac{1}{4}\sum_{h'_1, h'_2} s_{p_1 p_2 h'_1 h'_2} <h'_1\,h'_2\,|\,\bar{V}\,|\,h_1\, h_2>\nonumber\\
&&\hbox{with single-particle energy:}\,\epsilon_{k}=<k\,|\,T\,|\, k> +\sum_{h}<k\,h\,|\,\bar{V}\,|\,k\, h>\equiv <k\,|\,h_0\,|\, k>.
\end{eqnarray}
\end{adjustwidth}

The solution of this equation can be found as follows:
\begin{equation}
s_{p_1 p_2 h_1 h_2}=   \frac{<p_1\, p_2\,|\,\bar{G}(E=\epsilon_{h_1} +\epsilon_{h_2})\,|\,h_1\, h_2>}{\epsilon_{h_1} +\epsilon_{h_2}-\epsilon_{p_1}-\epsilon_{p_2 }}, \label{eq_spphh}
\end{equation}
where $G(E)$, the G matrix, is a two-body operator satisfying 
\begin{eqnarray}
 && G(E)=V\,+\, V \,\frac{Q}{E-h_0}\,G(E) \Longleftrightarrow{}\nonumber\\
&&  <k_1\, k_2\,|\,G(E)\,|\,k'_1\, k'_2>=<k_1\, k_2\,|\,V\,|\,k'_1\, k'_2> \nonumber\\
 && + \sum_{p_1, p_2 >k_F} <k_1\, k_2\,|\,V\,|\,p_1\, p_2>\frac{1}{E-\epsilon_{p_1}-\epsilon_{p_2 }}<\,p_1\, p_2\,|\,G(E)\,|\,k'_1\, k'_2>,
\end{eqnarray}
and $Q$ is the Pauli-blocking operator projecting on particle states above the Fermi sea. A different derivation of this result, not based on a variational approach, is given in Ref. \cite{Desplanques}. It follows that the binding energy of nuclear matter is 
\begin{equation}
		E_0=\sum_{h<k_F}<h\,|\,T\,|\, h> +\frac{1}{4} \sum_{h_1 h_2<k_F}<h_1\,h_2\,|\,\bar{G}(E=\epsilon_{h_1} +\epsilon_{h_2})\,|\,h_1\, h_2>,
  \end{equation}
  reflecting the fact that the dynamics of dressed HF nucleons is governed by the G matrix. The important point is that this $G$  matrix, often identified with the effective interaction in the Hartree--Fock calculation, refers to objects, the dressed or renormalized nucleons, which can be seen as Landau quasi-particles that are definitely different from bare nucleons. If an energy-independent mean field is assumed, then the single-particle energy difference entering, for example, Equations \eqref{eq_ah}, \eqref{eq_ap} and \eqref{eq_spphh}  involves only the kinetic energy difference between particle and hole states. 
  
 \subsection{ Response Function and Occupation Numbers}\label{sec2.2}
 Using the correspondence between bare and dressed nucleon operators (Equations (\ref{eq_ah}) and  (\ref{eq_ap})), an explicit form of  the genuine particle and hole spectral functions (i.e., related to the bare nucleons)  can be  obtained (see \cite{CEM2022}):  
 \begin{eqnarray}
		S^h_{k,k'}(E)&=&\sum_n <0|a^\dagger_k|n>< n|a_{k'}|0>\,
		\delta\left(E+E_n^{A-1}-E_0^A\right) = Z_k \,\delta_{kk'}\, \delta\left(E-\epsilon_k\right)\,\rho_k\nonumber\\ 
		&+&\frac{1}{2}\,\sum_{p_2 h_3 h_4}\frac{<k\, p_2\,|\,\bar{G}\,|\,h_3\, h_4>\,
			<k'\, p_2\,|\,\bar{G}\,|\,h_3\, h_4>}{\left(\epsilon_k +\epsilon_{p_2}-\epsilon_{h_3}-\epsilon_{h_4 }\right)\,
			\left(\epsilon_{k'} +\epsilon_{p_2}-\epsilon_{h_3}-\epsilon_{h_4}\right)}\nonumber\\ 
		&&\delta\left(E-\left[\epsilon_{h_3}+\epsilon_{h_4}-\epsilon_{p_2}\right]\right)\,\bar\rho_k\,\bar\rho_{k'}
		\label{Spectral3}
	\end{eqnarray}
	\begin{eqnarray}
		S^p_{k',k}(E)&=&\sum_n <0|a_{k'}|n>< n|a^\dagger_k|0>\,
		\delta\left(E-E_n^{A+1}+E_0^A\right)\,= Z_k \,\delta_{kk'} \,\delta\left(E-\epsilon_k\right)\,\bar\rho_{k}\nonumber\\ 
		&+&\frac{1}{2}\,\sum_{h_2 p_3 p_4}\frac{<p_3\, p_4\,|\,\bar{G}\,|\,h_2\,k'>
			<p_3\, p_4\,|\,\bar{G}\,|\,h_2\,k>}{\left(\epsilon_k +\epsilon_{h_2}-\epsilon_{p_3}-\epsilon_{p_4 }\right)\,
			\left(\epsilon_{k'} +\epsilon_{h_2}-\epsilon_{p_3}-\epsilon_{p_4}\right)}\nonumber\\
		&&\delta\left(E-\left[\epsilon_{p_3}+\epsilon_{p_4}-\epsilon_{h_2}\right]\right)\,
		\rho_k\,\rho_{k'}.
		\label{Spectral2} 
	\end{eqnarray}
	
 By integrating the spectral function, the occupation numbers are also obtained:
 \begin{eqnarray}
		n_h\equiv 1-\Delta n^h= &=& 1\,-\,\frac{1}{2}\,\sum_{h_2 p_3 p_4}\left|\frac{<p_3\, p_4\,|\,\bar{G}\,|\,h_2\, h>}{\epsilon_h +\epsilon_{h_2}-\epsilon_{p_3}-\epsilon_{p_4 }}\right|^2\\
		n_p\equiv \Delta n^p &=& \frac{1}{2}\,\sum_{p_2 h_3 h_4}\left|\frac {<p\, p_2\,|\,\bar{G}\,|\,h_3\, h_4>} {\epsilon_p +\epsilon_{p_2}-\epsilon_{h_3}-\epsilon_{h_4 }}\right|^2,
	\end{eqnarray}
	which is a well-known form, quoted, for instance, in Ref. \cite{Ramos91}, Equations (2.31, 2.32). 
 
 We can now obtain the response function to a probe (such as the vector boson related to charged current neutrino interaction), which transfers an energy-momentum to the nucleus 
 $(\omega,{\bf q})$ and couples to individual nucleons with time-independent operators: 
	$
	{\cal{O}}(j)= \tau_j^\pm, \,\,\, ( \bfvec{\sigma}_j \cdot \widehat{q} ) \, 
	\tau_j^\pm, \,\,\, 
	( \bfvec{\sigma}_j \times \widehat{q} )^i
	\, \tau_j^\pm$\endnote{The $ \sigma\tau$ operators are replaced by the usual 1/2 to 3/2 transition operators $ST$ in the case of coupling to the $\Delta$.}. This response is defined as 
	\begin{equation}
		\label{eq_def_resp}
		R(\omega,{\bf q})=\sum_{n}\,|< n|\sum_{j=1}^A\,{\cal{O}}(j)\, e^{i{\bf q}\cdot{\bf x}_j} |0 >|^2\,\delta(\omega-E_n + E_0), 
	\end{equation}
	where $|n>$ and $E_n$ are the eigenstates and eigenvalues of the full nuclear Hamiltonian. 
	In the factorization approximation, this response can be expressed in terms of the above spectral functions according to: 
	\begin{eqnarray}
		R(\omega,{\bf q})&=&\sum_{k'_1, k'_2,k_1,k_2} <k'_1|{\cal{O}}^\dagger  e^{-i{\bf q}\cdot{\bf x}}|k'_2><k_2|{\cal{O}} e^{i{\bf q}\cdot{\bf x}}|k_1>\nonumber\\
		&&\int_{-\infty}^{\varepsilon_F} dE_1 \, S^h_{k_1,k'_1}(E_1)\,\int_{\varepsilon_F}^{\infty} dE_2\,S^p_{k'_2,k_2}(E_2)\,\delta(\omega-E_2+E_1).
		\label{Facresp}
	\end{eqnarray}
	
	The indices $k$ ($k=({\bf k},s,t)$) stand for the basis of single-particle states, and  $a_k$ and $a^\dagger_k$ are the associated destruction and creation operators with discrete normalization $\left\{a_k,a^\dagger_{k'}\right\}=\delta_{kk'}$. The expression of the response function involves the known coupling of the external (electroweak) probe to the bare nucleon. The leading term (Equation (34) in \cite{CEM2022})  is the pure p-h response quenched by Z factors. Correlations generate two additional contributions, the so-called correlation piece (Equation (35) in \cite{CEM2022})
    and the so-called polarization piece (Equation (36) in \cite{CEM2022}) involving 2p-2h excitations at the origin of the enhancement of the neutrino--nucleus quasielastic cross-section seen by the MiniBooNE collaboration \cite{Katori2009,MiniBooNe2010}. In addition, notice that, as explained in detail in  \cite{CEM2022}, the delta resonance states can be incorporated as extra particle states in all parts of this approach.

    An excellent indicator of the role of the correlations is provided by the  sum rule: 
    \begin{equation}
		S({\bf q})=\frac{1}{A}\int d\omega\,R(\omega,{\bf q})=\frac{4}{A}\sum_{\bf k} n_{\bf k}\left(1-n_{\bf k +\bf q}\right).
		\label{sumrule}
	\end{equation}
	
\begin{figure}[H]
		\includegraphics[width=0.8\textwidth]{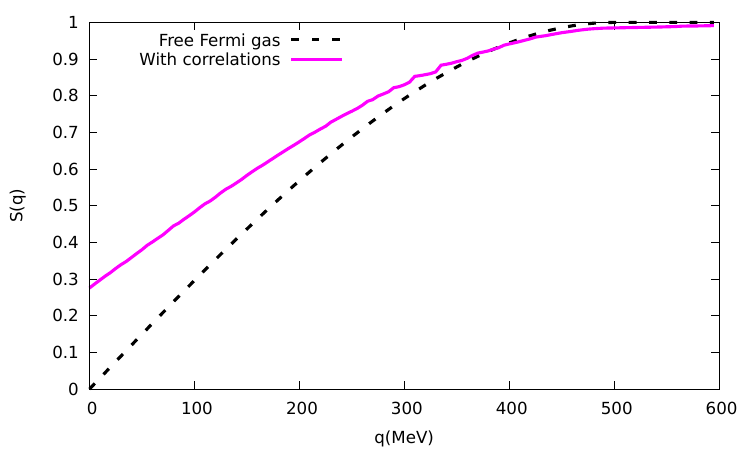}
		\caption{Sum rule as a function of momentum transfer in the free Fermi gas case and for including nucleon correlations. The value of the Fermi momentum is fixed at $k_F=245$ MeV.}
		\label{f1}
	\end{figure}

 In particular, at zero momentum, this sum rule does not vanish. Indeed, we established the result in \cite{CEM2022}:
 \begin{equation}
		S(0)\approx 2 \Delta n, 
\end{equation}
 where $\Delta n$ represents the mean depopulation (i.e., per nucleon) of the Fermi sea,
 \begin{equation}
		\Delta n=\frac{1}{2A}\,\sum_{h_1,h_2, p_1, p_2}\left|\frac{<p_1\, p_2\,|\,\bar{G}\,|\,h_1\, h_2>}{\epsilon_{h_1}+\epsilon_{h_2}-\epsilon_{p_1}-\epsilon_{p_2}}\right|^2 \equiv \kappa,\label{depop}
\end{equation}
 which just coincides with the dimensionless  wound integral of the Brueckner theory \cite{Tabakin,Machleidt89}, which is the norm of the defect wave function averaged over all NN pairs. The defect
wave function  is the difference between the correlated and uncorrelated NN wave functions,
and therefore, $\kappa$ is a measure of the importance of NN correlations, whose net effect is the depopulation of the Fermi sea. To show the effect of the correlation on the response function with a concrete example, we display in Figure \ref{f1} the result of the calculation (see \cite{CEM2022}) for the sum rule in a toy model adjusted to reproduce the values obtained in Ref.~\cite{Cioffi} for light nuclei such as carbon:
\begin{equation}
\Delta n=0.15,\qquad		<t>=\frac{3}{k^3_F}\int_0^\infty dk\, k^2\,\frac{k^2}{2M}\,n_{\bf k}=35\,MeV.   
\end{equation}
with $k_F=245\, MeV$ or $\rho=0.8\,\rho_0$. We can check that $S(0)\approx 2 \Delta n =0.3$ with good accuracy and that the enhancement due to the correlations persists up to a momentum $q=350\, \text{MeV}$.
 \subsection{The G Matrix and the Pair Correlation Function} \label{sec2.3}
To simplify the matter described below, we first omit the tensor force that we will reintroduce later perturbatively. We  separate the total CM momentum and the relative momentum of the two-nucleon states and write the G-matrix elements in each given spin--isospin ($S,T$) channel as
 \begin{eqnarray}
&&<\,\bfvec{k}_1\, \bfvec{k}_2\,|\,G^{ST}(E)\,|\,\bfvec{t}_1\, \bfvec{t}_2\,> = 
 \frac{1}{\mathcal{V}}\,<\,\bfvec{k}\,|\,G^{ST}_K(E)\,|\,\bfvec{t}\,>\,\delta_{\bf{K},\bf{T}}\equiv  \frac{1}{\mathcal{V} }\,G^{ST}_K(\bfvec{k}, \bfvec{t}; E)\,\delta_{\bf{K},\bf{T}}\\
&& \hbox{with:}\qquad \bfvec{k}_1=\frac{\bfvec{K}}{2}+\bfvec{k}, \quad\bfvec{k}_2=\frac{\bfvec{K}}{2}-\bfvec{k},\quad
 \bfvec{t}_1=\frac{\bfvec{T}}{2}+\bfvec{t},\quad\bfvec{t}_2=\frac{\bfvec{T}}{2}-\bfvec{t},
 \end{eqnarray}
 where all the plane-wave states are normalized to one in a box with a volume $\mathcal{V}$.
 In view of the nuclear matter calculation, the entry energy  will be the total energy of the two initial nucleons in the Fermi sea:
 \begin{equation}
E=\epsilon_{\bf{K}/2+\bf{t}} +\epsilon_{\bf{K}/2-\bf{t}}  < 2\epsilon_F .  
 \end{equation}
 
 Similarly, the matrix elements of the bare two-body potential are written as:
 \begin{eqnarray}
&&<\,\bfvec{k}_1\, \bfvec{k}_2\,|\,V^{ST}\,|\,\bfvec{t}_1\, \bfvec{t}_2\,> = 
 \frac{1}{\mathcal{V}}<\,\bfvec{k}\,|\,V^{ST}\,|\,\bfvec{t}\,>\,\delta_{\bf{K},\bf{T}}\\
&& <\,\bfvec{k}\,|\,V^{ST}\,|\,\bfvec{t}\,> \equiv V^{ST}(\bfvec{k}-\bfvec{t})=\int d\bfvec{r}\,e^{-i (\bf{k}-\bf{t})\cdot\bf{r}}\, V^{ST}(\bfvec{r}).
 \end{eqnarray}
 
 Just to fix the idea, let us first consider the case of a potential only  involving  the sigma, omega, pion and rho exchange channels.  In the simplest non-relativistic scheme, ignoring recoil correction, one has (with standard notations) for the most important spin--isospin states (i.e., the ones surviving in the  orbital $s$ state after antisymmetrization):
\begin{eqnarray}
 && S=0, T=1 \quad\hbox{and}\quad S=1,T=0\,\hbox{states}:\nonumber\\
 && V^{01}(q)=V_{\sigma}(q) +V_{\omega}(q)- \left(V_{\pi}(q) + 2\,V_{\rho}(q)\right)= V^{10}(q)\nonumber\\
 &&V_\sigma(q)= -g^2_\sigma\,\frac{1}{q^2+m^2_\sigma}\,\Gamma^2_S(q),\quad 
 V_\omega(q)= g^2_V\,\frac{1}{q^2+m^2_V}\,\Gamma^2_V(q),\nonumber\\
&& V_\pi(q)=-\left(\frac{g_A}{2 F_\pi}\right)^2\frac{q^2}{q^2+m^2_\pi}\,\Gamma^2_\pi(q),\quad
 V_\rho(q)=-\left(\frac{g_A}{2 F_\pi}\right)^2\,C_\rho\,\frac{q^2}{q^2+m^2_\rho}\,\Gamma^2_\rho(q),\quad \nonumber\\
 &&\hbox{with}\quad C_\rho=\left(\frac{F_\pi\,g_\rho\,(1+\kappa_\rho)}{g_A\,M_N}\right)^2,\qquad\Gamma_\alpha(q)=\frac{\Lambda^2_\alpha}{\Lambda^2_\alpha +q^2},\label{POTNN}
\end{eqnarray}
where $q$ represents the modulus of the exchanged three-vector momentum. The form factors that mathematically regularize the UV  (i.e., the short-distance) behavior of the potential are physically motivated by the compositeness and the finite size of the nucleon source of the emitted mesonic fields. In particular, they remove the contact piece of the central part of the (spin--isospin) $\pi$ and $\rho$ exchange and make the globally repulsive interaction at $r=0$ finite, transforming the hard core into a soft core. Of course, they do not affect the globally attractive long-range part of the interaction, dominated by the pion exchange and rho exchange in the strong rho scenario 
 ($\kappa_\rho\sim 6$) \cite{HP75}:
\vspace{-9pt}
\begin{adjustwidth}{-\extralength}{0cm}
\centering 
\begin{equation}
\left[V^{01}\right]_{long\,range}(r)=- g^2_S\,\frac{e^{-m_S r}}{4\pi r}\, +\, g^2_V\,\frac{e^{-m_V r}}{{4\pi r}}
 -\left(\frac{g_A}{2 F_\pi}\,\right)^2\left(m^2_\pi \frac{e^{-m_\pi r}}{4\pi r}\, +\,2\,C_\rho\, m^2_\rho \,\frac{e^{-m_\rho r}}{4\pi r}\right).
\end{equation}
\end{adjustwidth}

\begin{figure}[H]
		
		\includegraphics[width=0.8\textwidth]{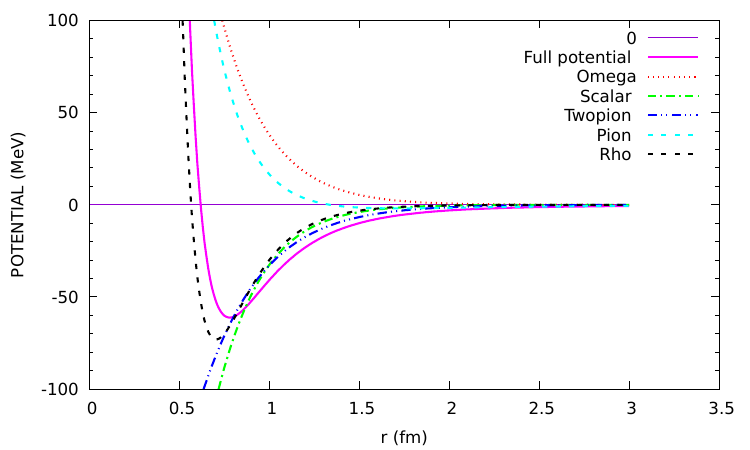}
		\caption{The full bare 
  $NN$  potential  (solid line) and the various contributions listed in Equation \eqref{POTNN}. The values of the parameters are given in Equation \eqref{CHIRPOT}.}
		\label{f2}
	\end{figure}

The combination of the repulsive short-range piece and the attractive long-range piece produces a pocket (see Figure \ref{f2}) able to generate an approximately zero-energy bound state. Of course, distinguishing the singlet channel ($S=0, T=1$) from the triplet deuteron channel ($S=1, T=0$) necessitates the introduction of the tensor potential related to the difference between pion and rho exchange potentials and the time component of the Lorentz vector part of the rho exchange, with these two effects making the triplet interaction significantly more attractive than the singlet interaction.

In order to isolate the role of the short-range correlations, we introduce a mixed representation of the G matrix  (see Ref. \cite{Baldo2012}) according to:
\begin{eqnarray}
 G^{ST}_K(\bfvec{r}, \bfvec{t})&=&<\,\bfvec{r}\,|\,G^{ST}_K(E)\,|\,\bfvec{t}\,>  =\int\, \frac{d \bfvec{q}}{(2\pi)^3}\,e^{i\bf{q}\cdot\bf{r}}\,G^{ST}_K(\bfvec{q}, \bfvec{t}; E)\\
 V^{ST}(\bfvec{r}, \bfvec{t})&=&<\,\bfvec{r}\,|\,V^{ST}\,|\,\bfvec{t}\,>  
 =\int \frac{d\bfvec{q}}{(2\pi)^3}\,e^{i\bf{q}\cdot\bf{r}}\, <\,\bfvec{q}\,|\,V^{ST}\,|\,\bfvec{t}\,>=V^{ST}(\bfvec{r})\,e^{i\bf{t}\cdot\bf{r}}.
\end{eqnarray}

 This suggests the need to search for a solution for the G matrix in the form:
\begin{equation}
G^{ST}_K(\bfvec{r}, \bfvec{t})=V^{ST}(\bfvec{r}) \,f^{ST}_K(\bfvec{r}, \bfvec{t}).   
\end{equation}

It follows that the G-matrix equation is equivalent to an equation for the correlation function $f^{ST}_K(\bfvec{r}, \bfvec{t}; E)$:
\begin{eqnarray}
 f^{ST}_K(\bfvec{r}, \bfvec{t}) &=& e^{i\bf{t}\cdot\bf{r}}\,+\,\int d\bfvec{y}\,\left[\frac{Q}{e}\right]_K(\bfvec{r}-\bfvec{y})\,V^{ST}(\bfvec{y})\,f^{ST}_K(\bfvec{y}, \bfvec{t})\\
\left[\frac{Q}{e}\right]_K(\bfvec{r}-\bfvec{y}) &=&
\int \frac{d\bfvec{q}}{(2\pi)^3}\,e^{i\bf{q}\cdot(\bf{r}-\bf{y})}\,\frac{P_{\bf{K}/2+\bf{q}}\,P_{\bf{K}/2 -\bf{q}} }{\epsilon_{\bf{K}/2+\bf{t}} +\epsilon_{\bf{K}/2-\bf{t}} - \epsilon_{\bf{K}/2+\bf{q}} - \epsilon_{\bf{K}/2 -\bf{q}}}.
\end{eqnarray}

 All these equations can be rewritten in an operator form,
\begin{eqnarray}
 G^{ST}_K(E)&=& V^{ST}\,F^{ST}_K(E),\quad   G^{ST}_K(E)=V^{ST}+V^{ST}\left[\frac{Q}{e}\right]_K (E)\,\,G^{ST}_K(E)\nonumber\\
F^{ST}_K(E)&=& 1 +\left[\frac{Q}{e}\right]_K (E)\,V^{ST}\,F^{ST}_K(E)=1+ \left[\frac{Q}{e}\right]_K (E)\,\,G^{ST}_K(E),
\end{eqnarray} 
 in terms of operators acting in the space of relative motion and defined by their matrix elements:
 \begin{eqnarray}
&&<\,\bfvec{r}\,|\,1\,|\,\bfvec{t}\,>=  e^{i\bf{t}\cdot\bf{r}},\quad
<\,\bfvec{r}\,|\,V^{ST}\,|\,\bfvec{t}\,>=  e^{i\bf{t}\cdot\bf{r}}\,V^{ST}(\bfvec{r}) \quad
<\,\bfvec{r}\,|\,F^{ST}_{K}(E)\,|\,\bfvec{t}\,>=f^{ST}_K(\bfvec{r}, \bfvec{t})\nonumber\\
&&<\,\bfvec{r}\,|\,G^{ST}_{K}(E)\,|\,\bfvec{t}\,>=G^{ST}_K(\bfvec{q}, \bfvec{t})=F_K(\bfvec{q}, \bfvec{t})\,V^{ST}(\bfvec{r}),\nonumber\\
&&<\,\bfvec{x}\,|\,\left[\frac{Q}{e}\right]_K (E)\,|\,\bfvec{y}\,>=
\left[\frac{Q}{e}\right]_K(\bfvec{x}-\bfvec{y}; E).
 \end{eqnarray}
 
 The correlation function can be expanded in partial waves, and one can define a correlation function for each two-body channel identified by the quantum numbers $STl$:
 \begin{eqnarray}
f^{ST}_K(\bfvec{r}, \bfvec{t})&=&\sum_{lm}\,4\pi \,(i)^l g^{STl}_K(r,t)\,Y_{lm}(\hat{\bfvec{r}}) \,Y^*_{lm}(\hat{\bfvec{t}}).   \\
\hbox{with}:&&\quad g^{STl}_K(r,t)\,\,\equiv\,\,f^{STl}_K(r,t)\,\,j_l(tr).
 \end{eqnarray}
 
 The resulting integral equation in the channel $STl$ reads:
 \begin{eqnarray}
 &&g^{STl}_K(r,t)=j_l(tr)+\int_0^\infty   \,4\pi y^2 dy\, \left[\frac{Q}{e}\right]_K(r,y)\,V(y)\,g^{STl}_K(y,t)\label{PARWAVE}\\
 &&\hbox{with}:\quad\left[\frac{Q}{e}\right]_K(r,y)=\int\frac{4\pi q^2  dq}{(2\pi)^3}\,j_l(qr)\,\bar{\left[\frac{Q}{e}\right]}_K(q)\,j_l(qy),\\ 
 &&\hbox{with}:\quad \bar{\left[\frac{Q}{e}\right]}_K(q)=
\int \frac{d\hat{\bfvec{q}}}{(4\pi)}\,\,\frac{P_{\bf{K}/2+\bf{q}}\,P_{\bf{K}/2 -\bf{q}} }{\epsilon_{\bf{K}/2+\bf{t}} +\epsilon_{\bf{K}/2-\bf{t}} - \epsilon_{\bf{K}/2+\bf{q}} - \epsilon_{\bf{K}/2 -\bf{q}}}.
 \end{eqnarray}
 
This equation is approximate in the sense that the Pauli-blocking operator divided by the energy denominator in relative momentum space  has been replaced by its angle average. It nevertheless becomes exact for zero CM momentum. The functions $f^{STl}_K(r,t)= g^{STl}_K(r,t)/j_l(tr)$  play the role of correlation functions to be compared with the state-independent correlation function $f(r)$ of the variational LOCV approaches (see~\cite{Baldo2012} and references therein).
 At variance with the variational framework,  these Brueckner correlation functions depend on the two-body channel and on the initial state with total CM momentum $K$ ($P$ in \cite{Baldo2012}) and relative momentum $t$ ($q$ in \cite{Baldo2012}). However, according to the results of Ref. \cite{Baldo2012}, for the dominant $l=0$  Fourier component at  normal nuclear density, one finds that the correlation functions for the $^1S_0\, (S=0, T=1)$ channel  and the $^3S_1 \,(S=1, T=0)$ channel (which is sensitive to the tensor force) calculated at given $t$ and $K$ values are very similar. In addition, it is found that these correlation functions are also very similar to the state-channel-independent correlation function of the variational LOCV  method. Moreover, using the $v18$ Argonne potential  \cite{Argonne}, this calculation demonstrates that the dependence on the CM momentum and on the relative momentum is also  weak. Hence, for convenience and practical reasons, but not only these reasons (see below), we propose representing the effect of short-range correlations by a unique state and channel-independent Jastrow function as:
 \begin{equation}
 f_\alpha(r)=1\,-\,\alpha\,j_0(q_c r) .\label{JASTROW}  
 \end{equation}
 
 Indeed, it was demonstrated in Ref. \cite{Nakayama} that, starting from the Bonn potential, this simple Jastrow ansatz can rather accurately reproduce the Brueckner reaction matrix in the dominant channels (but with state-dependent $q_c$ and $\alpha$ parameters). Of course, in the Brueckner theory, the correlation function has no reason to vanish  (i.e, $\alpha=1$) at the origin. For instance, one systematically has $f(r=0)\sim 0.2$ (or $\alpha=0.8$ for the corresponding Jastrow ansatz) for various values of  the relative momentum, of the CM momentum and of the density in the results reported in \cite{Baldo2012}. The fact that two nucleons may coexist at the same point is, mathematically, a consequence of modeling through the presence of the form factor transforming the hard core potential into a soft core potential; i.e., $V(r=0)$ remains finite.  One can very well consider that this modeling approach is not appropriate at very short distances if we state that two composite nucleons cannot coexist at the same place as a six-quark cluster, which has never been observed. For this reason, one can take the following ansatz  for the Jastrow function:
 \begin{equation}
 f^{STl}_K(r,t)=f_c(r)=1\,-\,j_0(q_c r)    
 \end{equation}
satisfying $f_c(r=0)=0$.
 An immediate consequence is that the effective interaction automatically vanishes at high momentum, providing the natural regularization of the loop integral entering, for instance, the two-pion (or two-rho) exchange contribution to the energy per nucleon of nuclear matter. As discussed in \cite{Chanfray1985}, this property can be seen as a consequence of the Beg--Agassi theorem, which states that virtual mesons propagate freely inside the correlation hole. However, the function $g^{01l}_K(r,t)=f_c(r)\,j_0(tr)$ cannot be an exact solution of the integral Equation (\ref{PARWAVE}). To find the optimal choice, we proceed as follows: we enter the function  $f_c(r)=1-j_0(q_cr)$ into the r.h.s of Equation (\ref{PARWAVE}) and calculate the output function $\Tilde{f}_c (r)$. We thus fix the value of $q_c$ to obtain $\Tilde{f}_c (r=0)=0$. Using typical parameters of the Bonn potential  (potential A in \cite{BONNPOT}), we obtain a value of the order of $q_c\sim 600$ MeV, which changes very little if we vary the input parameters. We show the result of the calculation for a Bonn-like non-relativistic OBE potential but with the parameters taken from or inspired by the chiral confining model in its QCD-connected version, as described below in Section \ref{sec3}:  
 \begin{eqnarray}
 &&g_\sigma=8.37,\quad m_\sigma=919\,\text{MeV},\quad \Lambda_S=1\,\text{GeV},\nonumber\\ 
 &&g_V=7.5,\quad m_V=783\,\text{MeV},\quad \Lambda_V=1\,\text{GeV},\nonumber\\
 &&g_A=1.26,\quad F_\pi=92.4\,\text{MeV},\quad m_\pi=140\,\text{MeV},\quad \Lambda_\pi=1\,\text{GeV},\nonumber\\
 &&g_\rho=g_V/3,\quad \kappa_\rho=6,\quad C_\rho=1.865 ,\quad m_\rho=770\,\text{MeV},\quad\Lambda_\rho=2\,\text{GeV}. \label{CHIRPOT}
 \end{eqnarray}
 
 In this version of the model, one actually introduces two scalar potentials: one associated with the QCD-connected scalar field, $s$, describing the fluctuations of the chiral condensate (see next section), and the other, $\sigma'$, simulating the two-pion (and two-rho) exchange potential with the delta resonance in the intermediate state. In the traditional OBE model, the correlated and uncorrelated two-pion exchanges  correspond to all the diagrams in
Figure 3 in Ref. \cite{MHE89} (apart from the iterative process with NN intermediate states). The parameters are: 
 \begin{equation}
g_{2\pi}=4.8\equiv g_{\sigma'},\quad m_{2\pi}=550\,\text{MeV}\equiv m_{\sigma'},\quad \Lambda_{2\pi}=1\,\text{GeV}\equiv \Lambda_{\sigma'}.\label{SIGPRIM}
 \end{equation}
 
\begin{figure}[H]
		
		\includegraphics[width=0.8\textwidth]{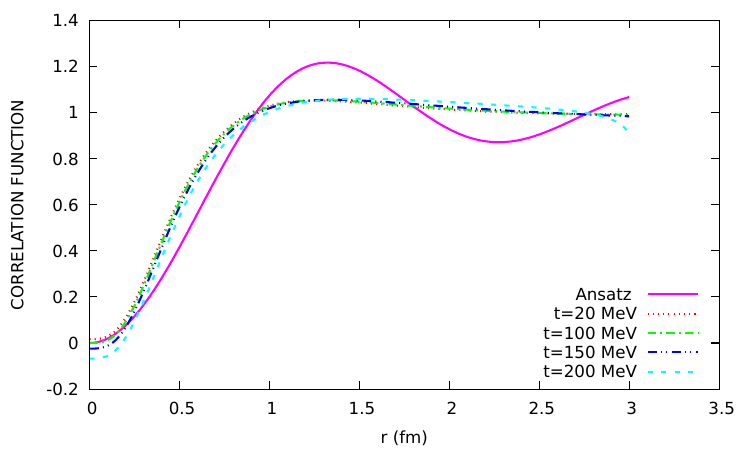}
		\caption{The input correlation 
 function $f_c(r)$ (solid line) and the output correlation  function after one iteration  using Equation \eqref{PARWAVE} for various values of the relative momentum $t$ and for zero total momentum $K$. The calculation is performed at normal nuclear matter density.}
		\label{f3}
	\end{figure}

We show the various contributions to this simple non-relativistic $NN$ potential in Figure~\ref{f2} and note the existence of an attractive pocket in the full potential, which is expected to be slightly deepened when the tensor force (needed to produce the deuteron bound state) and the time component of the rho exchange are incorporated.  We recall in passing that such a strong rho scenario (i.e., a strong (Lorentz) tensor coupling, $K_\rho= 6$) is required to decrease the (Wigner) tensor force without obtaining too large a D-state probability in the deuteron.\\
In Figure \ref{f3}, we show the output correlation function $\Tilde{f}(r)$ after one G-matrix iteration for a two-nucleon system in the $l=0$ state at  normal nuclear density for either the singlet channel, $^1S_0\, (S=0, T=1)$, or the triplet channel, $^3S_1 \,(S=1, T=0)$. We fix $q_c=670$ MeV to have a vanishing correlation function for a relative momentum $t=100$~MeV. As can be seen in this figure, varying $t$ induces a very moderate change in the correlation function, which is reminiscent of the very weak relative momentum dependence of the correlation function observed in Ref. \cite{Baldo2012}. In addition, varying $g_{\sigma'}$ and $m_{\sigma'}$  by $\sim$20\% or varying the density between zero  and twice the nuclear matter density  induces a change in $q_c$ of only a  few tens of MeV. 

If we now accept that the correlation function is well represented by this Jastrow function, the effective tensor force can be obtained very simply as $G_T(r)=V_T(r)f_c(r)$. For the total spin--isospin interaction written in momentum space, this explicitly gives: 
\begin{eqnarray}
G_{\sigma\tau}(\bfvec{q})&=&\int d\bfvec{r} \,e^{-i\bf{q}\cdot \bf{r}}\,  
\bigg[\int \frac{d\bfvec{k}}{(2\pi)^3}   \,e^{i\bf{k}\cdot \bf{r}}\bigg(\frac{1}{3}\big(V_\pi(k)+2V_\rho(k)\big)\,\bfvec{\sigma}_1 \cdot\bfvec{\sigma}_2 \nonumber\\ 
&&+\,\frac{1}{3}\,\big(V_\pi(k)-V_\rho(k)\big)\,\big(3\,\bfvec{\sigma}_1\cdot\hat{\bf{k}}\,\bfvec{\sigma}_2 \cdot\hat{\bf{k}}\, -\,\bfvec{\sigma}_1\cdot\bfvec{\sigma}_2\big)\bigg)\bigg]\,\, \left(1-j_0(q_c r)\right). 
\end{eqnarray}

One can adopt two different decompositions for the effective spin--isospin interaction: central and tensor or longitudinal and transverse,
\begin{eqnarray}
G_{\,\sigma\tau}(\bfvec{q})&=&\left(\frac{g_A}{2 F_\pi}\right)^2\,\left(v_c(q) \,\bfvec{\sigma}_1 \cdot\bfvec{\sigma}_2\,+\,v_t(q)\,\big(\bfvec{3\,\sigma}_1\cdot\hat{\bf{q}}\,\bfvec{\sigma}_2 \cdot\hat{\bf{q}}\, -\,\bfvec{\sigma}_2\cdot\bfvec{\sigma}_2\big)\right)\\
&=&\left(\frac{g_A}{2 F_\pi}\right)^2\,\left(v_L(q)\,\bfvec{\sigma}_1\cdot\hat{\bf{q}}\,\bfvec{\sigma}_2 \cdot\hat{\bf{q}}\,
\,+\,v_T(q)\,\bfvec{\sigma}_1 \times\hat{\bf{q}}\cdot\bfvec{\sigma}_2 \times\hat{\bf{q}}\right), \label{GEFFSI1}
\end{eqnarray}
with 
\begin{eqnarray}
v_c(q)&=& \frac{1}{3}\big(v_\pi(q) + 2v_\rho(q) + 3g'(q)\big),\qquad
v_t(q)= \frac{1}{3}\big(v_\pi(q) - v_\rho(q) + 3h'(q)\big)\nonumber\\
v_L(q)&=&v_\pi(q) + g'(q)+ 2h'(q),\qquad\qquad
v_T(q)=v_\rho(q) + g'(q)- h'(q), \label{GEFFSI2}
\end{eqnarray}
and
\begin{eqnarray}
v_\pi(q)&=&-\Gamma^2_\pi(q^2 )\frac{q^2}{q^2  +m^2_\pi}\nonumber\\       
v_\rho(q)&=&-C_\rho\,\Gamma^2_\rho(q^2 )\frac{q^2}{q^2  +m^2_\rho}\nonumber\\
g'(q)&=&-\frac{1}{3} \,\int\frac{d\bfvec{k}}{(2\pi)^3}\,\big(v_\pi(k)+ 2v_\rho(k)\big)\,\frac{2\pi^2}{q_c}\,\delta\left(|\bfvec{q}-\bfvec{k}|\,-\,q_c\right)\nonumber\\
&\simeq &\frac{1}{3}\,\Gamma^2_\pi(q^2 +q^2_c)\frac{q^2 +q^2_c}{q^2 +q^2_c +m^2_\pi}\,+\,\frac{2\,C_\rho}{3}\,\Gamma^2_\rho(q^2 +q^2_c)\frac{q^2 +q^2_c}{q^2 +q^2_c +m^2_\rho}\nonumber\\
h'(q)&=&-\frac{1}{3} \,\int\frac{d\bfvec{k}}{(2\pi)^3}\,\big(v_\pi(k) - v_\rho(k)\big)\,\,\big(3(\bfvec{k}\cdot\hat{\bf{q}})^2 -1\big)\,\frac{2\pi^2}{q_c}\,\delta\left(|\bfvec{q}-\bfvec{k}|\,-\,q_c\right)\nonumber\\
&\simeq &\frac{1}{3}\,\Gamma^2_\pi(q^2 +q^2_c)\frac{q^2}{q^2 +q^2_c +m^2_\pi}\,-\,\frac{C_\rho}{3}\,\Gamma^2_\rho(q^2)\frac{q^2}{q^2 +q^2_c +m^2_\rho}.
\label{GEFFSI3}
\end{eqnarray}

The analytical forms of $g'(q)$ and $h'(q)$, which are exact for vanishing and large momenta, provide an accurate interpolation between these two limits. This is the well-known result quoted in \cite{OTW}. In the Landau limit, we obtain the Landau--Migdal parameter $g'=g'(0)=0.59$, which is compatible with the range of accepted values \cite{Ichimura}. We can also remark that a large value of the cutoff for the $\rho NN$ (Lorentz) tensor coupling is needed to obtain a value for g' that is not too small. Finally, we also note that the whole G-matrix interaction is only made of terms in the form $F(q^2)-F(q^2 + q^2_c)$, which vanishes in the UV for $q>>q_C$ as a consequence of the Beg--Agassi--Gal theorem.

 It is well known that nuclear matter calculations based on a pure two-body interaction fail to simultaneously reproduce the binding  energy and the saturation density (the famous Coester band problem) even if relativistic effects (Walecka mechanism) \cite{SerotWalecka1986,Walecka1997}  included in the 
Dirac--Brueckner--Hartree--Fock (DBHF) approach may improve the situation, hence providing a guide for parameterization for an in-medium RMF with density-dependent coupling constants simulating many-body forces \cite{VanDalen}. In the following, we follow another route, where effects linked to the quark substructure and the broken chiral symmetry of the QCD vacuum generate the needed three-body forces. This is performed in the chiral confining model that we describe in the next section.

	\section{The Chiral Confining Model}\label{sec3}
\subsection{The Phenomenological Model}\label{sec3.1}
The very early motivation of this chiral model for dense  nuclear matter (and neutron star interior) was to find a firm theoretical basis of the very successful relativistic theories initiated by Walecka and collaborators \cite{SerotWalecka1986,Walecka1997}. This type of approach indeed provides a very economical saturation mechanism, and a spectacular well-known success is the correct magnitude of the spin--orbit potential since  the nucleons move in an 
attractive background scalar field and in a repulsive vector background field, which contribute in an additive way. Although the origin of the repulsive vector field can be safely identified as associated with the omega vector-meson exchange, the real nature of the attractive Lorentz scalar field has been  a controversial subject since there is no sharp scalar resonance that would lead to a simple scalar particle exchange. More fundamentally, the question of the very nature of these background fields has to be elucidated; in other words, it is highly desirable to clarify their relationship with the QCD condensates and, in particular, with the chiral quark condensate and, more generally, with the low-energy realization of chiral symmetry, which is spontaneously broken in the QCD vacuum and is expected to be progressively restored when the density increases.

To bridge the gap between relativistic theories of the Walecka type and approaches insisting on chiral symmetry, it was proposed in \cite{Chanfray2001} (see also the detailed discussion given in Ref. \cite{Martini2006}) that the ``nuclear physics scalar sigma meson'' of the relativistic Walecka model at the origin of the nuclear binding be identified; let us call it $\sigma_W$, with the chiral invariant $s=S-F_\pi$ field associated with the radial fluctuation of the chiral condensate $S$ around the ``chiral radius'' $F_\pi$, identified to the pion decay constant. Said differently, we take the point of view that the effective theory has to be formulated, as a starting point, in terms
of the fields associated with the fluctuations of the chiral quark condensate parameterized in a matrix form: 
\begin{eqnarray}
W&=&\sigma + i\vec{\tau}\cdot\vec{\pi}\equiv S\, U\equiv (s\, +\, F_{\pi})\,U\equiv (\sigma_W +\, F_{\pi})\,U\nonumber\\
&&\hbox{with}\qquad U(x)=e^{i\,{\vec{\tau}\cdot\vec{\phi}(x)}/{F_\pi}}.\label{REPRES}
\end{eqnarray}

The scalar field $\sigma$ ($S$) and pseudoscalar fields $\vec{\pi}$ ($\vec{\phi}$) written in Cartesian (polar) coordinates  appear as dynamical degrees of freedom and may deviate from the vacuum value, $\left\langle \sigma\right\rangle_{vac}=\left\langle S\right\rangle_{vac} = f_\pi\propto\left\langle \bar{q}q\right\rangle_{vac}$. The sigma and pion fields, associated with the amplitude  $s\equiv\sigma_W$ and phase fluctuations $\phi$ of this condensate, are promoted to the rank of effective degrees of freedom. Their dynamics are governed by an effective potential, $V_\chi\left(\sigma,\vec{\pi})\right)$, having a typical Mexican hat shape associated with a broken (chiral) symmetry of the QCD vacuum.

There is, however,  a well-identified problem concerning nuclear saturation with the usual chiral effective theories \cite{Boguta83,KM74,BT01,C03}. Independently of the particular chiral model, in the nuclear medium, one can move away from the minimum of the vacuum effective potential (Mexican hat potential), i.e., into a region of smaller curvature. This single-effect 
equivalent to the lowering of the sigma mass destroys the stability, creating problems for the applicability of such effective theories in the nuclear context. The effect can be associated with an $s^3$ tadpole diagram, generating attractive three-body forces and destroying saturation, even if the repulsive three-body force from the Walecka mechanism is present. One possible way to solve this problem is to introduce the nucleonic response to the scalar field, $\kappa_\mathrm{NS}$, which is the central ingredient of the quark--meson coupling model (QMC), introduced in the original pioneering work of P. Guichon \cite{Guichon1988} and successfully applied to finite nuclei
with an explicit connection to the Skyrme force \cite{Guichon2004,Guichon2004a,Guichon2004b}. This effect, which is associated with the polarization of the quark substructure in the presence of the nuclear scalar field, will unavoidably generate  three-body forces, which may provide the desired repulsion. In practice, this response or, more precisely, the nucleon scalar  susceptibility $\kappa_\mathrm{NS}$ generates a non-linear coupling of the scalar field to the nucleon or, equivalently, a decrease in the scalar coupling constant with increasing density. Hence, to achieve saturation, in a set of successive works \cite{Chanfray2005,Chanfray2007,Massot2008,Massot2009,Rahul}, we have complemented the relativistic chiral approach in such a way that the effect of the nucleon response is able to counterbalance the attractive chiral tadpole diagram to obtain good saturation properties, especially the correct curvature coefficient---the empirical incompressibility parameter.

Phenomenologically, the model is described with  standard notations by the Lagrangian
\begin{equation}
{\cal L}=\bar\Psi\,i\gamma^\mu\partial_\mu\Psi\,+\,
{\cal L}_s\,+\,{\cal L}_\omega\,+\,{\cal L}_\rho\,\,+\,{\cal L}_\pi,
\end{equation}
with
\begin{eqnarray}
{\cal L}_s &=& -M^*_N(s)\bar\Psi\Psi\,-\,V(s)\,+\,\frac{1}{2} \partial^\mu s\,\partial_\mu s\nonumber\\	
{\cal L}_\omega &=& - g_V\,\omega_\mu\,\bar\Psi\gamma^\mu\Psi\,+\,\frac{1}{2}\,m^2_V\,\omega^\mu\omega_\mu
\,-\,\frac{1}{4} \,F^{\mu\nu}F_{\mu\nu}\nonumber\\	
{\cal L}_\rho &=&- g_\rho\,\rho_{a\mu}\,\bar\Psi\gamma^\mu \tau_a\Psi
\,-\,g_\rho\frac{\kappa_\rho}{2\,M_N}\,\partial_\nu \rho_{a\mu}\,\Psi\bar\sigma^{\mu\nu}_{}\tau_a\Psi
\,+\,\frac{1}{2}\,m^2_\rho\,\rho_{a\mu}\rho^{\mu}_{a}
\,-\,\frac{1}{4} \,G_a^{\mu\nu}G_{a\mu\nu}\nonumber\\	
{\cal L}_\pi &=& \frac{g_A}{2\,F_\pi}\,\partial_\mu\varphi_{a\pi}\bar\Psi\gamma^\mu\gamma^5\tau_a\Psi
-\,\frac{1}{2}\,m^2_{\pi}\varphi_{a\pi}^2
\,+\,\frac{1}{2}\, \partial^\mu\varphi_{a\pi}\partial_\mu \varphi_{a\pi}.
\end{eqnarray}

It involves  the scalar field $s$ (the ``nuclear physics sigma meson'' $\sigma_W$), the pion field $\varphi_{a\pi}$, and the vector fields associated with the omega meson  channel ($\omega^\mu$) and with  the rho meson channel ($\rho_a^\mu$). Each meson-nucleon vertex is regularized by a meson-nucleon form factor mainly originating from the compositeness of the nucleon, hence generating a bare OBE Bonn-like NN interaction, as given in Equation \eqref{POTNN}. The specific crucial ingredients beyond the simplest approach are the presence of a chiral effective potential associated with the chirally broken vacuum and an in-medium-modified nucleon mass, which is supposed to embed its quark substructure.

In the majority of our previous works \cite{Chanfray2005,Martini2006,Chanfray2007,Massot2008,Massot2009,Massot2012,Rahul}, the chiral effective potential had the simplest linear sigma model (L$\sigma$M) form:
 \begin{equation}
V_{\chi,{L\sigma M}}(s)=\frac{1}{2}\,M^2_\sigma \,s^2\, +\,\frac{1}{2}\frac{M^2_\sigma -M^2_\pi}{ F_\pi}\, s^3\,+\,
\frac{1}{8}\,\frac{M^2_\sigma -M^2_\pi}{ F^2_\pi} \,s^4 \label{eq:VLSM}.
\end{equation}

The effective Dirac nucleon mass $M^*_N(s)$ deviates from the bare nucleon mass in the presence of the nuclear scalar field $s$:
\begin{eqnarray}
		M^*_N(s)&=& M_N + g_S\, s +   \frac{1}{2}\kappa_{NS} \,s^2 + \mathcal{O}(s^3). 
  \end{eqnarray}
where $g_S$ is the scalar coupling constant of the model. In \cite{Chanfray2005,Martini2006,Chanfray2007,Massot2008,Massot2009,Massot2012}, we took the pure linear sigma model value $g_S=M_N/F_\pi$, but in the most recent work \cite{Rahul}, $g_S$ was allowed to deviate from L$\sigma$M and was fixed by performing a Bayesian analysis. This quantity actually corresponds to the first-order response of the nucleon to an external scalar field and can be obtained in an underlying microscopic model of the nucleon. The nucleon scalar  susceptibility $\kappa_\mathrm{NS}$ is another response parameter and reflects the polarization of the nucleon, i.e., the self-consistent readjustment of the quark wave function in the presence of the scalar field.  Very generally, the scalar coupling constant, $g_S$, and the nucleon response parameter,  $\kappa_{NS}$, depend on the subquark structure and the confinement mechanism, as well as the effect of spontaneous chiral symmetry breaking.
In our previous works \cite{Chanfray2005,Chanfray2007,Massot2008,Massot2009,Massot2012,Rahul}, we introduced a dimensionless parameter, 
\begin{equation}
 C\equiv \frac{\kappa_\mathrm{NS}\,F_\pi^2}{2 M_N},   
\end{equation}
which is expected to be of the order $C\sim 0.5$ as in the MIT bag  used in the QMC framework. The physical motivation to introduce this nucleonic response is the observation that nucleons experience huge fields at finite density, e.g., the scalar field is of the order of a few hundred MeV at saturation density. Nucleons, being composite objects in reality, will react to the nuclear environment (i.e., the background nuclear scalar field) through the (self-consistent) modification of quark wave functions. 
This effect may generate a three-body force that provides the desired repulsion 
if confinement dominates spontaneous chiral symmetry breaking in the nucleon mass origin, as discussed in Ref. \cite{Chanfray2011} within particular models. Indeed, it is possible to show that the scalar sector generates a three-nucleon repulsive contribution to the energy per nucleon:
\begin{equation}
\frac{E^{(3)}}{A}\simeq\frac{g^2_S}{2\,M^4_\sigma}\,\left(\kappa_\mathrm{NS}-\frac{g_S}{F_\pi}\right)\,\rho^2_s\,
=\,\frac{g^3_S}{2\,M^4_\sigma\,F_\pi}\,\left(2\, \Tilde{C}_3\,-\,1\right)\,\rho^2_s\qquad
\hbox{with}\qquad \Tilde{C}_3 \simeq\frac{M_N}{g_S\,F_\pi}\,C .
\label{eq:THREEBOD}
\end{equation}

This is the result previously quoted in Equation~(44) in Ref.~\cite{Chanfray2007}, but without the factor $M_N/g_S F_\pi$, which appears when the scalar coupling constant is allowed to deviate from its pure L$\sigma$M value, as in Ref. \cite{Rahul}. This three-body effect provides a very natural mechanism for the saturation mechanism but is at variance with the QMC model, which ignores the attractive tadpole diagram present in the chiral approach, this requires a $C$ parameter that is close to or even larger than one \cite{Chanfray2005,Chanfray2007,Massot2008,Massot2009,Massot2012,Rahul}.

One very important point  is that it is possible to relate these scalar response parameters to the chiral properties of the nucleon, namely, the first derivative of the nucleon mass with respect to to the current quark mass (or to the pion mass), which is related to the light-quark sigma commutator and the second derivative, linked to the chiral susceptibility of the nucleon.  The nucleon mass and other intrinsic properties of the nucleon (sigma term, chiral susceptibilities) are indeed QCD quantities, which are, in principle, obtainable from lattice simulations.   The problem is that lattice calculations of this kind are difficult for small current quark mass, $m $, or equivalently, small  pion mass, $M^2_\pi$.
Here, $M_\pi$ represents the pion mass to the leading order in the quark mass (i.e., ignoring the NLO chiral logarithm correction), $M^2_\pi=2 m \,B=-2 m\,\langle\overline{q}\,q\rangle_{\chi L}/{F^2}$ (Gell--Mann--Oakes--Renner (GOR) relation). The quantities  $B$ and $F$, the pion decay constant in the chiral limit, are two low-energy parameters appearing in chiral perturbation theory \cite{Leut2012}. The difficulty of the extrapolation is linked to the nonanalytical behavior of the nucleon mass as a function of $m$ (or equivalently, $M^2_\pi$), which comes from the pion cloud contribution. The idea of the Adelaide group \cite{LTY03,LTY04,TGLY04,AALTY10} was to separate the pion cloud self-energy, $\Sigma_{\pi}(M^2_{\pi}, \Lambda)$, from the rest of the nucleon mass and to calculate it with just  one adjustable cutoff parameter $\Lambda$ entering the $\pi NN, \pi N\Delta$ form factor regularizing the pion loops. Actually, different cutoff forms for the pion loops
(Gaussian, dipole, monopole, sharp) were used  with the adjustable parameter $\Lambda$. This formulation of ChiPT is thus called the Finite-Range Regulator (FFR) method.  The remaining nonpionic part is expanded in terms of powers of $M^2_{\pi}$  as follows: 
\begin{equation}
    M_N(M_\pi^2) = a_0 + a_2 M_\pi^2 + a_4 M_\pi^4  + \dots + \Sigma_\pi(M_{\pi},\, \Lambda),
    \label{eq:lattice_1}
\end{equation}

Depending of the details of the chiral extrapolation, the extracted values of the parameters are within the range  of  $a_2\simeq 1.5 \,\text{GeV}^{-1},\,a_4\simeq -0.5 \,\text{GeV}^{-3}$ \cite{LTY04} and 
 $a_2\simeq 1.0\,\text{GeV}^{-1},\,a_4\simeq -0.25\, \text{GeV}^{-3}$ \cite{AALTY10}. 
The explicit connection between lattice QCD  parameters $a_2$ and $a_4$ and the response parameters $g_S$ and $C$ in the L$\sigma$M case was obtained in~\cite{Chanfray2007,Massot2008}: 
\begin{equation}
a_2= \frac{F_\pi\, g_{S}}{M^2_{\sigma}}, 
\qquad
a_4 =-\frac{F_\pi\,g_{S}}{2 M^4_{\sigma }}\,\left(3\,-\,2\,\tilde{C}_3 \right).\label{LATTRES}
\end{equation}

Notice that in the expression of $a_4$, the factor $M_N/F_\pi g_s$ in $\tilde{C}_3$, present in our recent paper \cite{Rahul}, was absent in \cite{Chanfray2007,Massot2008} since the nucleon mass was fixed to be $M_N=F_\pi g_s$.
The quantity $a_2 M^2_\pi \sim$ 20--30 MeV represents the nonpionic piece of the sigma commutator directly associated with the scalar field, $s$ (see the detailed discussion of this quantity in Ref.~\cite{Chanfray2007}).
One very robust conclusion  is that the lattice result for $a_4$ is much smaller than that obtained in the simplest linear sigma model, ignoring the nucleonic response ($C=0$), for which $a_4\simeq 3.5\,\text{GeV}^{-3}$. 
 Hence, lattice data require a strong compensation from the effects governing the three-body repulsive force needed for the saturation mechanism: compare Equations \eqref{LATTRES} and \eqref{eq:THREEBOD}.
Hence, both lattice data constraints and nuclear matter phenomenology require quite a large value of the dimensionless response parameter, $C$, which must be at least larger than one. Moreover, in a recent work based on a Bayesian analysis with lattice data as an input \cite{Rahul}, we found that the response parameter is strongly constrained to the value $C\sim 1.4$, very close to the value where the scalar susceptibility changes in sign: $C=1.5$.

\subsection{The NJL Chiral Confining Model}\label{sec3.2}
The problem that one has to face is that it seems impossible to find a realistic confining model for the nucleon able to generate $C$ larger than one. For instance, as mentioned above, in the MIT bag model used in the QMC scheme, one has $C_\mathrm{MIT}\simeq 0.5$. One possible reason for this discrepancy between models and phenomenological values of $C$ lies in the  use of the linear sigma model (L$\sigma$M), which  is probably too naive. Hence, one should certainly use an enriched chiral effective potential from a model able to give a correct description of the low-energy realization of chiral symmetry in the hadronic world.
A good, easily tractable candidate is the Nambu--Jona--Lasinio (NJL) model. Indeed, in Ref.~\cite{Chanfray2011}, an explicit construction of the background scalar field was performed in the NJL model using a bosonization technique based on an improved derivative expansion valid at low (space-like) momenta~\cite{Chan}. Various confining interactions have been incorporated (quark--diquark string interaction, linear and quadratic confining interaction) on top of the NJL model, which seem to be sufficient to generate saturation, although the response parameter $C$ remains relatively small, of the order of $C\sim  0.5$. The reason is that, for a given scalar mass, the NJL chiral effective potential generates a significantly smaller $s^3$ cubic term (hence generating a smaller attractive tadpole diagram) than the simplistic linear sigma model. We  discuss this point below  in more detail  and demonstrate that the repulsive three-body force generating saturation 
is determined not only by the nucleon response $C$ but also by the cubic term of the NJL potential, hereafter governed by the parameter $C_\chi$, with the parameters $C$ and $C_\chi$ combining together in the three-body interaction. Below, we briefly outline this approach, which is described in detail in a parallel paper \cite{Chanfray2023}.

Let us now  consider the NJL model defined by the Lagrangian:
\begin{eqnarray}
{\cal L}&=& \overline{\psi}\left(i\,\gamma^{\mu}\partial_\mu\,-\,m\right)\,\psi\,+\,\frac{G_1}{2}\,\left[\left(\overline{\psi}\psi\right)^2\,+\
\left(\overline{\psi}\,i\gamma_5\vec\tau\,\psi\right)^2\right]\nonumber\\
& &-\,\frac{G_2}{2}\,\left[\left(\overline{\psi}\,\gamma^\mu\vec\tau\,\psi\right)^2\,+\,
\left(\overline{\psi}\,\gamma^\mu\gamma_5\vec\tau\,\psi\right)^2\,+\,\left(\overline{\psi}\,\gamma^\mu\,\psi\right)^2\right]. \label{LNJL}
\end{eqnarray}
It depends on four parameters: the coupling constants $G_1$ (scalar) and $G_2$ (vector), the current quark mass $m$ and a (noncovariant) cutoff parameter $\Lambda$. Three of these parameters ($G_1$, $m$ and $\Lambda$) are adjusted to reproduce the pion mass, the pion decay constant and the quark condensate.  We refer the reader to \cite{Chanfray2011} for more details. 
Using path integral techniques (i.e., integrating out $q\bar{q}$ pairs in the Dirac sea) and after the chiral rotation of the quark field, it can be equivalently  written in a semi-bosonized form involving a pion field $\vec{\phi}$ embedded in the unitary operator $U=\xi^2=exp(i\,\vec{\tau}\cdot\vec{\phi}(x)/{F_\pi})$, a scalar field, ${\cal S}$, a vector field, ${V}^\mu$, and an axial-vector field,  ${A}^\mu$, with all these meson fields being  coupled to the constituent valence quarks. It has the explicit form given in Equations (2) and (7)--(11) in Ref.~\cite{Chanfray2011}. Subtracting vacuum expectation values, the chiral effective potential can be expressed as:
\begin{equation}
V_{\chi,\mathrm{NJL}}(s)=-2N_c N_f\,\big(I_0(\mathcal{S})\,-\,I_0(M_0)\big) \,+\,\frac{\left(\mathcal{S}-m\right)^2 -\left(M_0 - m\right)^2}{2\,G_1}.\label{eq:VNJL}
\end{equation}

The quantity $-2N_c N_f\,I_0(\mathcal{S})$ is nothing but the total (in-medium) energy of the Dirac sea of constituent quarks with the NJL loop integral $I_0(\mathcal{S})$ given in Ref. \cite{Chanfray2023}. The  $\mathcal{S}$ fields whose vacuum expectation value coincides with the vacuum constituent quark mass $M_0$ is related to the ``nuclear physics sigma meson'' field $s$ according to:
\begin{equation}
\mathcal{S} \equiv\frac{M_0}{F_\pi}\,\left(s+F_\pi\right).
\end{equation}

For a comparison with the usual RMF model using the L$\sigma$M chiral effective potentials in Equation~\eqref{eq:VLSM} or, equivalently, non-linear sigma couplings, we expand the effective potential to the third order in $s$ as:
\begin{equation}
V_{\chi,\mathrm{NJL}}({s})=V_{\chi}(0)+V'_{\chi}(0)\,{s}+\frac{1}{2}V''_{\chi}(0)\, {s}^2 +\frac{1}{6}V'''_{\chi}(0)\, {s}^3 +.....
\end{equation}
An explicit calculation of the derivatives of the potential yields \cite{Chanfray2023}:
\begin{equation}
V_{\chi,\mathrm{NJL}}(s)= \frac{1}{2}\,M^2_\sigma\, {s}^2\, +\,\frac{1}{2}\,\frac{M^2_\sigma -M^2_\pi}{ F_\pi}\, {s}^3\,\big(1\,-\,C_{\chi,\mathrm{NJL}}\big) +...,
\label{eq:vchiNJL}
\end{equation}

\begin{figure}[H]

\includegraphics[width=0.8\textwidth,angle=0]{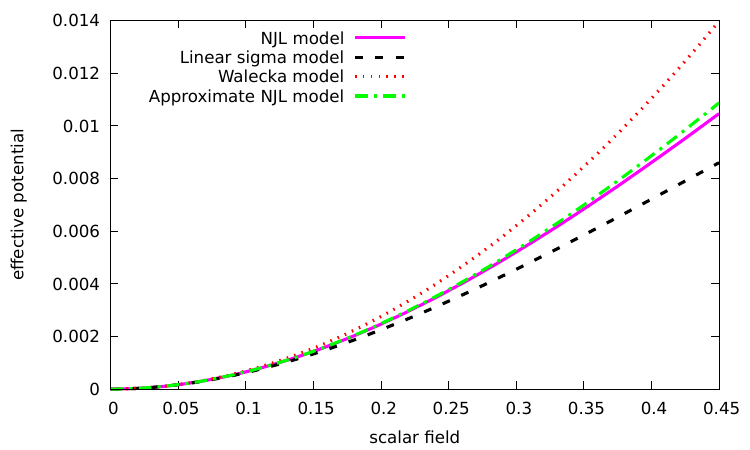}
\caption{Effective potential (in units of the string tension $\sigma^2$, with $\sigma=0.18$~GeV$^2$) plotted against $\vert s\vert/F_\pi$ for the NJL model (solid line), L$\sigma$M (dashed line) and original Walecka model limited to the quadratic term (dotted line). The figure also shows the approximate form of the NJL potential when limited to the cubic term in the scalar field $s$ expansion~\eqref{eq:vchiNJL} (dot-dashed line).  The values of the parameters are $G_1=12.514$, $G_2=0$, $\Lambda =0.604$~GeV and $m=5.8$~MeV, which yields $F_\pi=91.9$~MeV, $M_\pi=140$~MeV, $M_\sigma=716.4$~MeV  and $C_{\chi,\mathrm{NJL}}=0.488$. Note that the approximate expansion~\eqref{eq:vchiNJL}  is almost identical to the NJL potential.}
\label{fig:f2}
\end{figure}

The effective sigma mass $M_\sigma \sim 2 M_0$ and the pion mass $M_\pi$ and $F_\pi$ are calculated within the model. The quantity $C_{\chi,\mathrm{NJL}}$ is another NJL parameter, whose expression in terms of the NJL loop integral is given in \cite{Chanfray2023}. For a large cutoff value, the $C_{\chi,\mathrm{NJL}}$ parameter goes to zero, but for typical values of the  NJL parameters, its value is in the range 0.4--0.5. Figure~\ref{fig:f2} shows that the approximate expansion~\eqref{eq:vchiNJL} reproduces the exact NJL potential ~\eqref{eq:VNJL} very well. In this figure, the values of the NJL parameters are taken from the QCD-connected model presented in the next subsection. Comparing L$\sigma$M with the NJL scalar potential, one sees that the attractive tadpole term is larger in the case of L$\sigma$M for the same value of the effective sigma mass, $M_\sigma$. The effect of the parameter $C_{\chi,\mathrm{NJL}}$ is then the reduction in the attractive tadpole diagram and the greater repulsion of the chiral potential. In the same figure, we also show the potential of the  original Walecka model (again for the same effective sigma mass) that is limited to the quadratic term. 
As derived in \cite{Chanfray2023}, to the leading order in density, the scalar sector  generates a three-nucleon contribution contribution to the energy per nucleon, which is modified according to  (compare with Equation \eqref{eq:THREEBOD}):
\begin{eqnarray}
&&\frac{E^{(3b)}}{A}\simeq\frac{g^2_S}{2\,M^4_\sigma}\,\left(\kappa_\mathrm{NS}\,-\,\frac{g_S\,V'''_{\chi}(0)}{3\,M^2_\sigma}\right)\,\rho^2_s\,
=\,\frac{g^3_S}{2\,M^4_\sigma\,F_\pi}\,\left(2\, \Tilde{C}_3\,-\,1\right)\,\rho^2_s\nonumber\\
&&\hbox{with}\qquad \Tilde{C}_3 \simeq\frac{M_N}{g_S\,F_\pi}\,C +\,\frac{1}{2}\,C_\chi.
\label{eq:THREEBODN}
\end{eqnarray}

One very important point is that, as established in \cite{Chanfray2023}, the relationship between lattice QCD parameters is also modified according to (compare with Equation \eqref{LATTRES}):
\begin{equation}
a_2= \frac{F_\pi\, g_{S}}{M^2_{\sigma}}, \qquad  
a_4 =-\frac{F_\pi\,g_{S}}{2 M^4_{\sigma }}\,\left(3\,-\,2\,\tilde{C}_L \right)\quad\hbox{with}\quad
\tilde{C}_L = \frac{M_N}{g_S\,F_\pi}\,C\,+\,\frac{3}{2}\,C_\chi.\label{eq:LATTIX}
\end{equation}

The important conclusion discussed in detail in Ref. \cite{Chanfray2023} is that for a given three-nucleon force allowing saturation, one needs a lower value of the dimensionless parameter $C$ to obtain a small value of  the lattice parameter $a_4$ compatible with lattice data. To provide insight into  the effect of an enriched chiral effective potential, we return to our our original paper \cite{Chanfray2005}. In this paper, where L$\sigma$M was used ($g_S=M_N/F_\pi$), we  obtained the correct saturation properties with $C=1$  (see Figure~\ref{fig:f2} in this paper). We can retrospectively calculate the $a_2$ and $a_4$ parameters: we find $a_2=1.67$~GeV$^{-1}$ and $a_4=-1.48$~GeV$^{-3}$. If the value obtained for  $a_2$ is not very far from the lattice values, $a_4$ is at a magnitude three times larger than the upper value compatible with the lattice calculation. We can now play a little game by just incorporating the NJL-like $(1-C_\chi)$ correction in the cubic term term of the L$\sigma$M chiral  effective potential with $C_\chi=0.44$. Keeping all the other parameters at their original value, we take $C=0.78$ so as  to keep the same value of the repulsive three-body force, i.e., $ \Tilde{C}_3=C+C_\chi/2=1$  \eqref{eq:THREEBODN}. The new saturation point remains very close to the original one, but the $a_4$ parameter becomes very close to zero, $a_4=-0.1$~GeV$^{-3}$, in much better agreement with the lattice data mentioned above, namely,  $a_4\simeq -0.5 \,\text{GeV}^{-3}$ \cite{LTY04} and $a_4\simeq -0.25\, \text{GeV}^{-3}$ \cite{AALTY10}. See also the discussion given in \cite{Chanfray2023}. The conclusion is that the use of an enriched chiral effective potential of the NJL type significantly increases the agreement with lattice data,  together with the expected  model values of the nucleonic response parameter $C$. In the last section, we present the calculation of the nuclear matter equation of state with parameters taken from the QCD-connected model. We show explicitly that the inclusion of the $C_\chi$ parameter will generate a correct value of the $a_4$ parameter, together with a moderate value of the $C$ parameter, while preserving the saturation properties.

\subsection{The QCD-Connected Chiral Confining Model}\label{sec3.3}

The general picture underlying our approach can be summarized as follows. Nuclear matter is made of nucleons, which are themselves built from quarks and gluons and look like Y-shaped strings  generated by a nonperturbative confining force, with constituent quarks at the ends. These quarks acquire a large mass from a quark condensate, which is the order parameter associated with the spontaneous breaking of chiral symmetry in the QCD vacuum. When the density $\rho$ of nuclear matter increases, the QCD vacuum is modified by the presence of the nucleons, yielding a decrease in the quark condensate, indicating the progressive restoration of chiral symmetry. Hence, what is usually called ``the nuclear medium'' can be seen as a ``shifted vacuum'' with a lower value of the order parameter. The mass of the constituent quarks
coincides  with the in-medium expectation value, $M=\bar{\mathcal{S}}(\rho)$, of the chiral-invariant scalar field $\mathcal{S}$, associated with the radial fluctuation mode of the chiral condensate. 
To implement such a physical picture,  in this section, we  propose an effective model of low-energy QCD that incorporates both chiral symmetry breaking, i.e., the condensation of quark--antiquark pairs in the QCD vacuum,  and a confining string force that binds the massive constituent quarks inside the nucleon. It is based on the field correlator method (FCM) developed by Y. Simonov and collaborators \cite{Simonov1997, Simonov1998,Tjon2000,Simonov2002a,Simonov2002,Simonov-light, light1,light2,Digiacomo}, where the starting point is the Euclidean QCD Lagrangian and partition function written with conventional notation:
\begin {eqnarray}
L_{\QCD} &=&  L_0 + L_1+ L_G, \qquad  Z = \int d\psi\, d\bar\psi\, dA_\mu \,e^{-\left[L_0 + L_1+ L_G\right]},\nonumber\\                                     \hbox{with}\quad L_0 &=&\int d^4x\,\bar\psi(x)\left(\gamma_\mu\partial_\mu +m\right)\psi(x),\nonumber\\
L_1 &=& \int d^4x\,\bar\psi(x)\, ig\gamma_\mu A_\mu \,\psi(x), \qquad
L_G  =  -\frac{1}{4}\int d^4x\, F^a_{\mu\nu}F^a_{\mu\nu}.
\end{eqnarray}

In the  first step, we  integrate over the colored gluon fields ($A_\mu\equiv A^a_\mu\,t_a$) to generate a pure quark effective action. This is the gluon field averaging briefly described below. However, before carrying out this gluon integration, one makes a gauge choice (modified Fock--Schwinger gauge \cite{Simonov1997, Simonov1998,Tjon2000,Simonov2002a,Simonov2002,Simonov-light,light1,light2}) such that 
\begin{equation}
A_\mu\left(x_4,{\bf r}_0\right)=0,\,\,\,\,\left({\bf x}-{\bf r}_0\right)_j\cdot A_j\left(x_4,{\bf x}\right)=0,
\end{equation}
where ${\bf r}_0$ is an arbitrary point, which will be identified later with a string junction coordinate. In this particular gauge, it is possible  to express $A_\mu$ in terms of the field strength tensor $F_{\mu\nu}$~(\cite{Simonov1997, Simonov1998,Tjon2000,Simonov2002a,Simonov2002,Simonov-light,light1,light2}):
\begin{eqnarray}
A_4\left(x_4,{\bf x}\right)&=&\int_0^1 dv\,\left({\bf x}-{\bf r}_0\right)_k\, F_{k4}\left(x_4, {\bf r}_0+ v \left({\bf x}-{\bf r}_0\right)\right)\nonumber\\
A_j\left(x_4,{\bf x}\right)&=&\int_0^1 dv\,v\,\left({\bf x}-{\bf r}_0\right)_k\, F_{kj}\left(x_4, {\bf x}_0+ v \left({\bf x}-{\bf r}_0\right)\right).
\end{eqnarray}

Hence, the vector field appears as a contour integral, with the contour $C({\bf x})$ reducing to a straight line ${\bf z}(v)={\bf r}_0+ v \left({\bf x}-{\bf r}_0\right)$ between the $x_4$ axis and the point where the gluon field is calculated. The partition function can be written in successive forms,
\begin{eqnarray}
Z &=&\int d\psi\, d\bar\psi\, dA_\mu e^{-\left[L_0 + L_1+ L_G\right]}\equiv \int d\psi\, d\bar\psi\,e^{-L_0}\left\langle e^{-L_1}\right\rangle_G \nonumber\\
&\equiv & \int d\psi\, d\bar\psi\,e^{-\left[L_0 +L_1^{\eff}\right]}\equiv
\int d\psi\, d\bar\psi\,e^{-L^{\eff}},
\end{eqnarray}
where, in the second form, gluon field averaging has been performed, which subsequently defines the effective quark Lagrangian $L^{\eff}$ once the gluon field has been integrated out. In the field correlator method, the latter quantity can be obtained as an expansion of the exponential $\left\langle e^{-L_1}\right\rangle_G$. Using cluster expansion \cite{Simonov1997, Simonov1998,Tjon2000,Simonov2002}, $L^{\eff}$ can be written as an infinite sum containing averages such as $\langle(A_\mu)^k\rangle$. According to Ref. \cite{Simonov2002}, one can make a Gaussian approximation, neglecting all correlators 
$\langle(A_\mu)^k\rangle$ of degrees higher than $k=2$. The numerical accuracy has been studied on the lattice, demonstrating that the corrections to the Gaussian approximation are not larger than $3\%$ \cite{Simonov2002,Shev2000}.
Hence, neglecting all higher correlators, this effective Lagrangian (or more precisely, this effective action) reads
\begin{eqnarray}
L^{\eff}&\simeq& L_0 +\frac{1}{2}\left\langle L_1 L_1 \right\rangle_G = \int  d^4x\,\bar\psi(x)\left(\gamma_\mu\partial_\mu \,+\, m\right)\psi(x)\nonumber\\
&& +\frac{1}{2}\int d^4x\,d^4y\,\,\,\bar\psi(x)\, \gamma_\mu\, t_a \,\psi(x)\,\,\,\,\bar\psi(y)\, \gamma_\nu\, t_b\, \psi(y)
\,\,\,\,\frac{\delta_{ab}}{C_F}\,J^{\mu\nu}(x,y),
\end{eqnarray}
with $C_F=(N^2_c -1)/2 N_c=4/3$ and
\begin{eqnarray}
&&J^{\mu\nu}(x,y)=\frac{g^2}{2N_c}\left\langle A_a^\mu(x) A_a^\nu(y)\right\rangle\nonumber\\
&=&\int_0^1 dv\,\int_0^1 dw\,\alpha_\mu(v)\,\alpha_\nu(w)\,\left({\bf x}-{\bf r}_0\right)_k\,\left({\bf y}-{\bf r}_0\right)_q\,\frac{g^2}{2N_c}\left\langle F_a^{k\mu}\left(z(v)\right)\,F_a^{q\nu}\left( z'(w)\right)\right\rangle ,
\end{eqnarray}
with $z_4=x_4$, ${\bf z}(v)={\bf r}_0+ v \left({\bf x}-{\bf r}_0\right)$,
$z'_4=y_4$, ${\bf z}'(w)={\bf r}_0+ w \left({\bf y}-{\bf r}_0\right)$, $\alpha_4(v)=1$, 
$\alpha_k(v)=v$.
This expression of the kernel can be shown to be gauge-invariant, the reason being that the parallel transporters on the contour $C(x,y)$ are identically equal to unity in this gauge. The whole nonperturbative physics is contained in the nonlocal gluon condensate $g^2\left\langle F_a^{k\mu}\left(z(v)\right)\,F_a^{q\nu}\left( z'(w)\right)\right\rangle$, and parameterization is known from lattice measurements \cite{Digiacomo}. As in many previous works by Simonov and collaborators, we keep only the nonperturbative confining piece,
\begin{equation}
\frac{g^2}{2N_c}\left\langle F_a^{\rho\mu}(z)\,F_a^{\lambda\nu}(z')\right\rangle=\left(\delta_{\rho\lambda}\delta_{\mu\nu}-\delta_{\rho\nu}\delta_{\mu\lambda}\right) D(z-z'),
\end{equation}
where $D(x)$ decreases in all directions and describes the profile of the bilocal correlator of the nonperturbative gluonic fields in the QCD vacuum. It depends on two QCD quantities, the gluon condensate $\mathcal{G}_2\sim 0.015$~GeV$^4$ and the gluon correlation length $T_g$, which  physically corresponds to the string width. It may also be parameterized in terms of two QCD parameters, the string tension $\sigma=0.18\,GeV^2$ and $T_g$. We adopt a convenient Gaussian parameterization approach \cite{Tjon2000,Simonov2002}: 
\begin{equation}
D(x)=D(0) \, e^{-x^2/4 T^2_g}\,\,\,\,\hbox{with}\,\,\,\,\,D(0)=\frac{\sigma}{2\pi T^2_g}=\frac{\pi^2}{18}\,\mathcal{G}_2,\,\,\,\,\hbox{hence }\,\,\,\,\,\sigma=\frac{\pi^3}{9}\,\mathcal{G}_2\,T^2_g.
\end{equation}

 This form  of the bilocal correlator, which has the advantage of simplifying the calculations,  was justified in \cite{Tjon2000,Simonov2002}. In short, since all the observables are integrals of $D(x)$, its explicit form is not essential at large
distances, provided that it has a finite range $T_g$ and  that the string tension, i.e., the coefficient in the area law of the Wilson loop, is equal to $\sigma=\int d^2u\, D(u)/2$. Numerically, the gluon correlation length can be estimated as:
\begin{equation}
T_g=0.336 \,\left(\frac{\sigma\,((\mathrm{GeV}^2)}{0.2}\right)^{1/2}\left(\frac{0.02}{\mathcal{G}_2\,(\mathrm{GeV}^4)}\right)^{1/2}\, \mathrm{fm}.
\end{equation}
A ``small'' parameter is: 
\begin{equation}
\eta^2= \sigma\,T^2_g =
0.2\,\left(\frac{\sigma\,(\mathrm{GeV}^2)}{0.2}\right)\left(\frac{T_g\,(\mathrm{fm})}{0.1973}\right)^{2}\,=\,0.462\,\left(\frac{\sigma\,(\mathrm{GeV}^2)}{0.2}\right)\left(\frac{T_g\,(\mathrm{fm})}{0.3}\right)^{2}
.
\end{equation}

As in Ref. \cite{Tjon2000,Simonov2002}, we will ignore the magnetic piece of the kernel, keeping only the dominant  electric piece $J_{44}$. In addition, we make a static approximation, ignoring retardation effects:
\begin{equation}
e^{-x_4^2/4 T^2_g}\simeq 2\sqrt{\pi}\, T_g \,\delta (x_4).
\end{equation}

This approximation, already made in Ref. \cite{BBRV98} in the context of the heavy--light quark system, can be valid if the energy scale $T_g^{-1} \sim 700-800$ MeV is bigger than the other scale $\sigma\sim 400$ MeV of the problem or, equivalently, if the parameter $\eta$ is smaller than one.  Hence, we end up with an effective static Lagrangian and, consequently, a more tractable effective Hamiltonian governing the dynamics of light quarks in the presence of  a three-quark junction (or a static heavy quark) placed in ${\bf r}_0$ in the QCD vacuum. Said differently, this effective Hamiltonian has to be seen as a  way to derive baryonic Green's function and  bound-state equations \cite{Alkofer2000,Eich2016,Alkofer2019,Jakovac2019}, already given in Refs. \cite{Simonov1997, Simonov1998,Simonov2002}. In principle, the point ${\bf r}_0$ is arbitrary and should be chosen as the one minimizing the string lengths (Torricelli point) joining  the three quarks (see the discussion after Equation (28) in \cite{Simonov2002}). In practice, we take it as a constant parameter coinciding with the three-quark string junction, and the three-quark wave function is simply expressed in a factorized form in terms of single-quark orbitals centered in ${\bf r}_0$ \cite{Simonov2002}. Moreover, as discussed below,  a strong point of this approach is that vacuum chiral symmetry breaking will show up as being intimately related to the confinement mechanism. Physically, this means that the three-quark string keeping the quarks together is constructed on top of a chirally broken vacuum, building a constituent quark at the end of the strings \cite{BBRV98}. Indeed, the effective  Hamiltonian reads: 
\vspace{-9pt}
\begin{adjustwidth}{-\extralength}{0cm}
\centering 
\begin{equation}
H=\int d^3x\,\psi^\dagger({\bf x})\left(-i\,\vec{\alpha}\cdot\vec{\nabla}\,+\, m\right)\psi({\bf x})
\,+\,\frac{1}{2}\int d^3x\,\int d^3y\,\psi^\dagger({\bf x})\,t_a\,\psi({\bf x})\,V\left({\bf x},{\bf y}\right)\,
\psi^\dagger({\bf y})\,t_a\,\psi({\bf y}).
\end{equation}
\end{adjustwidth}

Introducing  ${\bf R}=({\bf x+y})/2-{\bf r}_0$, ${\bf r}={\bf x-y}$, ${\bf X}={\bf x}-{\bf r}_0$ and ${\bf Y}={\bf y}-{\bf r}_0$, the potential $V$, derived from $J_{44}$ \cite{Tjon2000,Simonov2002}, can be written in various forms:
\begin{eqnarray}
V\left({\bf x},{\bf y}\right)&=&\frac{1}{C_F}\,\frac{\sigma}{\sqrt{\pi}\,T_g}\,\left({\bf x}-{\bf r}_0\right)\cdot\left({\bf y}-{\bf r}_0\right)\,
\int_0^1 dv\,\int_0^1 dw\,e^{-\frac{\left(v\left({\bf x}-{\bf r}_0\right)-w\left({\bf y}-{\bf r}_0\right)\right)^2}{4 T^2_g}}\equiv \nonumber\\
V\left({\bf R},{\bf r}\right)&=&\frac{1}{C_F}\,\frac{\sigma}{\sqrt{\pi}\,T_g}\,\left({\bf R}^2-\frac{{\bf r}^2}{4}\right)\,\int_0^1 dv\,\int_0^1 dw\,e^{-\left(\frac{\left(v+w\right)\bf r}{4T_g}+
\frac{\left(v-w\right){\bf R}}{2T_g}\right)^2}\equiv\nonumber\\
V\left({\bf X},{\bf Y}\,;\,{\bf r}\right)&=&\frac{1}{C_F}\,\frac{\sigma}{\sqrt{\pi}\,T_g}\,\left(\frac{{\bf X}^2 +{\bf Y}^2}{2}-\frac{{\bf r}^2}{2}\right)\,I\left({\bf X},{\bf Y}\,;\,{\bf r}\right)\nonumber\\
\hbox{with}&&\quad I\left({\bf X},{\bf Y}\,;\,{\bf r}\right)\,=\,\int_0^1 dv\,\int_0^1 dw\,e^{-\left(\frac{\left(v+w\right)\bf r}{4T_g}+
\frac{\left(v-w\right)}{2T_g}\frac{{\bf X}+{\bf Y}}{2}\right)^2}.
\label{kernel}
\end{eqnarray}

The relative variable ${\bf r}$ is the one associated with the self-interaction of the quark. Keeping only the ${\bf r}$ dependence in the second form of Equation \eqref{kernel}, namely, taking ${\bf R}=0$, i.e, ${\bf r}_0$ at the mid-point between ${\bf x}$ and ${\bf y}$, the $V({\bf r})$ interaction allows the generation of a BCS-like chirally broken vacuum and a mass gap equation for the constituent quark. The light $q \bar{q}$ mesons and, in particular, the  Goldstone pion mode can be generated using this approach in a way equivalent to the RPA in the many-body nuclear problem \cite{Cotanch}. It is also equivalent to the quark Dyson--Schwinger (gap) and bound-state Bethe--Salpeter equations~\cite{Alkofer2000,Alkofer2007,Alkofer1989}. In the FCM approach, this program is implemented using a Fierz transformation and a  bosonization procedure, taking the origin ${\bf r}_0$ at the mid-point between the quark and the antiquark \cite{Simonov-light,light1,light2}; see also Section VII-D of Ref. \cite{Simonov2019} for a short summary of this method. 
The variable ${\bf R}$  physically corresponds to the length of the confining string. Indeed, at a large value of $R$, one can analytically check that the integral of the Gaussian factor in Equation~\eqref{kernel} has a $1/R$ behavior, which, together with the $R^2$ prefactor, yields a linear behavior: $C_F V(R,r)\sim \sigma R$.  We see (Equation~\eqref{kernel}) that, in practice, the self-interacting part and the string-like confining piece of the kernel might be mixed up in a very complicated way in the case of the presence of an ``external'' source, either a string junction or a static heavy (anti)quark. The treatment of  chiral symmetry breaking versus confinement entanglement is certainly the most prominent problem of QCD, which already manifests  at the level of the bare nucleon, as well as at the level of dense and hot nuclear or hadronic matter. However, some approximation schemes have been developed in the literature to partially disentangle these two aspects \cite{BBRV98,KNR2017}. Inspired by the third writing of the potential in Equation (\ref{kernel}), we propose a specific ansatz interpolating between the ansatz used in Refs. \cite{BBRV98,KNR2017}:
\begin{equation}
V\left({\bf X},{\bf Y}\,;\,{\bf r}\right)= \tilde{V}_C({\bf X})\,+\,\tilde{V}_C({\bf Y})\,+\,\tilde{V}_{CSB}({\bf r}),    
\end{equation}
with 
\vspace{-9pt}
\begin{adjustwidth}{-\extralength}{0cm}
\centering 
\begin{equation}
\tilde{V}_{CSB}({\bf r})=-\frac{1}{C_F}\,\frac{\sigma}{2\,\sqrt{\pi}\,T_g}\,{\bf r}^2
\,I\left(0\,,0\,;\,{\bf r}\right)=-\frac{1}{C_F}\,\frac{\sigma}{2\,\sqrt{\pi}\,T_g}\,{\bf r}^2
\int_0^1 dv\,\int_0^1 dw\,e^{-\left(\left(v+w\right)^2\frac{{\bf r}^2}{16 T^2_g}\right)}
\end{equation}
\end{adjustwidth}
\begin{equation}
\tilde{V}_C({\bf X})=\frac{1}{C_F}\,\frac{\sigma}{2\,\sqrt{\pi}\,T_g}\,{\bf X}^2
\,I\left({\bf X},{\bf X}\,;\,0\right)=\frac{1}{C_F}\,\frac{\sigma}{2\,\sqrt{\pi}\,T_g}\,{\bf X}^2
\,\int_0^1 dv\,\int_0^1 dw\,e^{-\left(\left(v-w\right)^2\frac{{\bf X}^2}{4 T^2_g}\right)}
\end{equation}
\begin{equation}
\tilde{V}_C({\bf Y})=\frac{1}{C_F}\,\frac{\sigma}{2\,\sqrt{\pi}\,T_g}\,{\bf Y}^2
\,I\left({\bf Y},{\bf Y}\,;\,0\right)=\frac{1}{C_F}\,\frac{\sigma}{2\,\sqrt{\pi}\,T_g}\,{\bf Y}^2
\,\int_0^1 dv\,\int_0^1 dw\,e^{-\left(\left(v-w\right)^2\frac{{\bf Y}^2}{4 T^2_g}\right)}.
\end{equation}

It turns out that $\tilde{V}_{CSB}({\bf r})$ has the following form:
\begin{equation}\
\tilde{V}_{CSB}({\bf r})=\frac{1}{C_F}\,\frac{4\sigma\,T_g}{\sqrt{\pi}}\left(v_l(r/T_g)-1\right),
\end{equation}
where $v_l(r)$ is a rapidly  decreasing function of $r$ with $v_l(0)=1$ and 
$v_l(\infty)=0$. This suggests decomposing the potential  according to
\begin{equation}
V\left({\bf X},{\bf Y}\,;\,{\bf r}\right)= V_C({\bf X})\,+\,V_C({\bf Y})\,+\,V_{CSB}({\bf r}),  \label{STRUCT}    
\end{equation}
with
\vspace{-9pt}
\begin{adjustwidth}{-\extralength}{0cm}
\centering 
\begin{equation}
V_{CSB}\left({\bf r}\right)=\frac{1}{C_F}\,\frac{2\,\sigma\,T_g}{\sqrt{\pi}}\,\left(2\,-\,\frac{{\bf r}^2}{4\,T^2_g}\,\int_0^1 dv\,\int_0^1 dw\,e^{-\left(\left(v+w\right)^2\frac{{\bf r}^2}{16 T^2_g}\right)}\right)\equiv\frac{1}{C_F}\,\frac{4\,\sigma\,T_g}{\sqrt{\pi}}\,v_l(r/T_g)
\end{equation}
\begin{equation}
V_C\left({\bf R}\right)=\tilde{V}_C({\bf R})-\frac{1}{C_F}\,\frac{2\,\sigma\,T_g}{\sqrt{\pi}}=\frac{1}{C_F}\,\frac{\sigma\,T_g}{2\,\sqrt{\pi}}\,\frac{{\bf R}^2}{T^2_g}\,\int_0^1 dv\,\int_0^1 dw\,e^{-\left(\left(v-w\right)^2\frac{{\bf R}^2}{4 T^2_g}\right)}-\frac{1}{C_F}\,\frac{2\,\sigma\,T_g}{\sqrt{\pi}}.\label{VCONF} 
\end{equation}
\end{adjustwidth}

Hence, the effective interaction Hamiltonian can be split according to: 
\begin{eqnarray}
H_{int}&=&H_{CSB}\,+\,H_C\label{HINT}\\
H_{CSB}&=&\frac{1}{2}\int d^3x\,\int d^3y\,\psi^\dagger({\bf x})\,t_a\,\psi({\bf x})\,V_{CSB}\left({\bf r}\right)\,
\psi^\dagger({\bf y})\,t_a\,\psi({\bf y})\label{HCHI}\\
H_C&=&\frac{1}{2}\int d^3x\,\int d^3y\,\psi^\dagger({\bf x})\,t_a\,\psi({\bf x})\,K_C\left({\bf X}, {\bf Y}\right)\,
\psi^\dagger({\bf y})\,t_a\,\psi({\bf y})\nonumber\\
&&\hbox{with}\quad K_C\left({\bf X}, {\bf Y}\right)=V_C\left({\bf X}\right)\,+\,V_C\left({\bf Y}\right)\label{HCONF}.
\end{eqnarray}

$H_{CSB}$ represents a short-range  interaction generating chiral symmetry breaking and the condensation of  light $q \bar{q}$ pairs in the QCD vacuum,  whereas $H_C$ represents a long-range interaction confining quarks inside the nucleon. In the following, the QCD vacuum and the ``shifted'' vacuum at finite density will be determined variationally as a BCS-like state $\left|\varphi(\rho)\right\rangle$. It is a simple matter to check that $\langle\varphi(\rho)|\,H_C\,|\varphi(\rho)\rangle =0$. Hence, the confining interaction in the presence of a string junction does not contribute to the chiral effective potential. Since this confining interaction exists only in the presence of an external source, i.e., a string junction, we can consider it to act only on valence quarks and not on the Dirac sea quarks.  The same mechanism  was described in Ref. \cite{BBRV98} for the heavy--light quark system, with a heavy quark seen as an external source located in ${\bf r}_0$.

\subsubsection{The Self-Energy Kernel}
\begin{figure}[H]
 \includegraphics[width=0.8\textwidth]{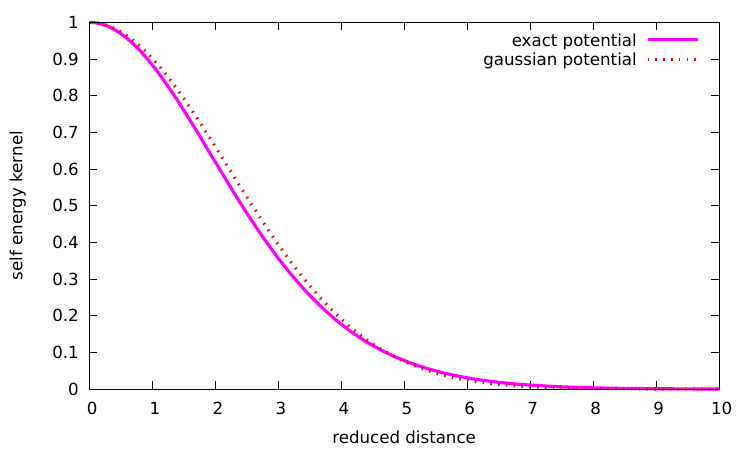}
  \caption{ The reduced self-energy kernel $v_l(r)$ normalized to $v_l(0)=1$ as a function of the reduced distance, $r/T_g$. The exact result (solid line) and the Gaussian approximation (dashed line) are shown (see text).}
  \label{SELF}
\end{figure}

The short-range ${\bf r}$-dependent piece $H_{CSB}$ will generate dynamical chiral symmetry breaking.  One numerically finds that $\int d^{3}r\, v_l(r/T_g)=\pi^{3/2}\, T^3_g\, \mu^{3}$ with $\mu^{3}=30$.
In Figure~\ref{SELF},  we compare the exact form with the approximate Gaussian form:
\begin{equation}
 v^G_l(r/T_g)=e^{-\frac{r^2}{\mu^2\,T^2_g}}= e^{-\left(\frac{r}{3.1\,T_g}\right)^2}.  
\end{equation}

This interaction can be rewritten as:
\vspace{-6pt}
\begin{adjustwidth}{-\extralength}{0cm}
\centering 
\begin{eqnarray}
 V_{CSB}\left({\bf r}\right)&=&\frac{1}{C_F}\,\frac{4\,\sigma\,T_g}{\sqrt{\pi}}\pi^{3/2}\, T^3_g\, \mu^{3}\,\frac{v_l(r/T_g)}{\int d^{3}r'\, v_l(r'/T_g)}=
 \frac{1}{C_F}\,4\pi\,\sigma\,T^4_g\,\mu^3\,\frac{v_l(r/T_g)}{\int d^{3}r'\, v_l(r'/T_g)}\nonumber\\
 &\simeq & \frac{1}{C_F}\,4\,N_c\,N_f\,G_1\,\delta({\bf r})\qquad\hbox{with}\qquad 4\,N_c\,N_f\,G_1\equiv 4\pi\,\sigma\,T^4_g\,\mu^3 \label{NJLR}. 
\end{eqnarray}
\end{adjustwidth}

In the second line of Equation (\ref{NJLR}), we have approximated the kernel by a point-like interaction with the same total strength. By performing a Fierz transform \cite{Simonov2002a,Simonov-light,,light1,light2} of the Lagrangian or Hamiltonian of this  type, $(\psi^\dagger t_a\psi )(\psi^\dagger t_a\psi )$, one can check that this interaction is equivalent to a Nambu--Jona--Lasinio (NJL) model  with a scalar sector coupling constant $G_1$ (we take $N_f=2,\, N_c=3)$:
\begin{equation}
G_1=\frac{4\pi\,\sigma\,T^4_g\,\mu^3}{4\,N_c\,N_f} = 3.1416\,  \left(\frac{\sigma\,(\mathrm{GeV}^2)}{0.2}\right)\left(\frac{T_g\,(\mathrm{fm})}{0.1973}\right)^{4}\, \mathrm{GeV}^{-2}.
\end{equation}

The reason for introducing a point-like interaction  is mainly practical. Keeping the finite-range CSB interaction would imply the use of bilocal meson  fields  resulting from the bosonization procedure, which was used in our previous  paper \cite{Chanfray2011}.  This cumbersome formalism  would have required completely reformulating and extending  our approach and would also lack its simplicity.  Conversely, one can also say that this FCM  approach justifies, on firm grounds, the NJL phenomenology, which is itself an important result, despite the approximations performed. The momentum space representation of this interaction can be written as: 
\vspace{12pt}
\begin{eqnarray}
\hat{V}_{CSB}\left({\bf q}\right)&=&\frac{1}{C_F}\,4\pi\,\sigma\,T^4_g\,\mu^3\,\Gamma ({\bf q})=\frac{1}{C_F}\,4\,N_c\,N_f\,G_1\Gamma ({\bf q})\nonumber\\
\Gamma ({\bf q})&=&\frac{\int d^{3}r\,exp\left(-i{\bf q}\cdot{\bf r}\right)\, v_l(r/T_g)}{\int d^{3}r\, v_l(r/T_g)}
\simeq exp\left(-\frac{\mu^2\,T^2_g\,q^2}{4}\right)\simeq \Theta\left(\Lambda\,-\,q)\right)\label{NJLQ}.
\end{eqnarray}

In the second line of Equation (\ref{NJLQ}), we have approximated the interaction in momentum space by the Gaussian form and then by a theta function with the cutoff $\Lambda$ which regularizes the equivalent NJL model. Since this  correspondence is not exact, fixing the equivalent NJL cutoff is somewhat arbitrary. A possible prescription is:
\begin{equation}
V_{CSB}\left(r=0\right)=\frac{1}{C_F}\,\frac{4\sigma\,T_g}{\sqrt{\pi}}\,=\int\frac{d^3 q}{(2\pi)^3}\,\hat{V}_{CSB}\left({\bf q}\right)
\simeq \frac{1}{C_F}\,4\pi\,\sigma\,T^4_g\,\mu^3\,\int\frac{d^3 q}{(2\pi)^3}\,\Theta\left(\Lambda\,-\,q\right).
\end{equation}

This yields:
\begin{equation}
 \Lambda = \left(\frac{6\sqrt{\pi}}{\mu^3}\right)^{1/3} \,\frac{1}{T_g}=0.71\,\left(\frac{0.1973}{T_g(\mathrm{fm})}\right)\, \mathrm{GeV}.
\end{equation}

One relevant dimensionless parameter to appreciate  the strength of the NJL equivalent interaction is: 
\begin{equation}
G_1\,\Lambda^2=\frac{(36\pi^4\mu^3)^{1/3}}{N_c N_f}\,\sigma T^2_g =7.86\,\sigma T^2_g =7.86\,\eta^2.   
\end{equation}

By considering the NJL gap equation \cite{Chanfray2023} in the chiral limit, $$1=24\,G_1\int_0^{\Lambda} \frac{d^3 p}{(2\pi)^3}\frac{1}{\sqrt{p^2+M_0^2}},$$ it is easy to establish that one must have $G_1 >\pi^2/3\simeq 3.29$ to have a nontrivial solution of the gap equation, i.e., a nonvanishing constituent quark mass, $M_0$. In practice, this necessitates a sizable $\eta^2 >0.5$.

Starting with accepted values of the string tension and gluon condensate, for orientation, we take $\sigma=0.18\,\mathrm{GeV}^2$ and  $\mathcal{G}_2=0.025\,\mathrm{GeV}^4$, which implies that $T_g= 0.286\,\mathrm{fm}$, and hence, $\eta=\sqrt{\sigma}T_g=0.615$. Consequently,  the NJL parameters are $G_1=12.514 \,\mathrm{GeV}^{-2}$ and $\Lambda =0.488\, \text{GeV}$. It follows that $G_1\Lambda^2=3$ is just below the critical value for chiral symmetry breaking. Hence, as in many QCD-inspired approaches, our model, at this level of approximation, is not able to properly generate the physical broken vacuum. We thus  decided to  increase the cutoff parameter to $\Lambda =0.604$ MeV to obtain the correct phenomenology.

We solve the NJL gap equation, $V'_{\chi,NJL}(\mathcal{S})=0$, with a light-quark mass $m=5.8$ MeV to obtain the vacuum constituent quark mass $\mathcal{S}=M_0$. With this set of parameters, we ignore the vector interaction and, consequently, the $\pi$-axial mixing  and calculate the pion decay constant, the pion mass and the quark condensate from the GOR relation, with all formal details being given in \cite{Chanfray2011}. We obtain the following results: $M_0=356.7\,\text{MeV},\, F_\pi=91.9\,\text{MeV},\,m_\pi=140\,\text{MeV}, \, \langle \Bar{q} q\rangle=-\left(241.1\,\text{MeV}\right)^3$. It is remarkable that by taking accepted values of the two basic QCD parameters, namely, the string tension and the gluon condensate, one recovers the NJL parameters, yielding good NJL phenomenology, or at least their order of magnitude. 

We also find that the NJL parameter entering the cubic term of the chiral effective potential is significant, $C_\chi =0.488$, and the effective sigma mass parameter is $M_\sigma=716.4$~MeV. We will show below  with the FCM confining potential that the scalar coupling constant (ignoring the effect of the pion cloud) is $g_S=6.52$, and the response parameter is $C=0.32$. 

\subsubsection{The Confining Kernel}
The confining interaction is given by Equation (\ref{VCONF}). We see immediately that it has a quadratic behavior at a short distance ($R<<T_g$), whereas at a large distance, it is possible to show that it has a linear behavior: 
\begin{eqnarray}
R<<T_g & :&  \qquad (V_C)_S\left({\bf\ R}\right)+\frac{1}{C_F}\,\frac{2\,\sigma\,T_g}{\sqrt{\pi}}=\frac{1}{C_F}\frac{\sigma}{2\,\sqrt{\pi}\,T_g}\, R^2\nonumber\\
R>>T_g & :&  \qquad (V_C)_L\left({\bf R}\right)+\frac{1}{C_F}\,\frac{2\,\sigma\,T_g}{\sqrt{\pi}}=\frac{1}{C_F} \,\sigma\,\left(R\,-\,\frac{2}{\sqrt{\pi}}\,T_g\right).
\end{eqnarray}

For the bound-state calculation,  to simplify the numerical computation, we make use of an approximate expression interpolating between the short-range and long-range behaviors. 
\begin{eqnarray}
 &&(V_C)_I\left({\bf R} \right) =F(R)\,(V_C)_S\left({\bf R}\right)\,+\,\left(1-F(R)\right)(V_C)_L\left({\bf R}\right).\nonumber\\
 &&\hbox{with}\qquad  F(R)=\left[1\,+\,exp\left(\frac{R-(6/\sqrt{\pi})\,T_g}{0.1\,T_g}\right)\right]^{-1} .
\end{eqnarray}

\begin{figure}[H]

  \includegraphics[width=0.8\textwidth,angle=0]{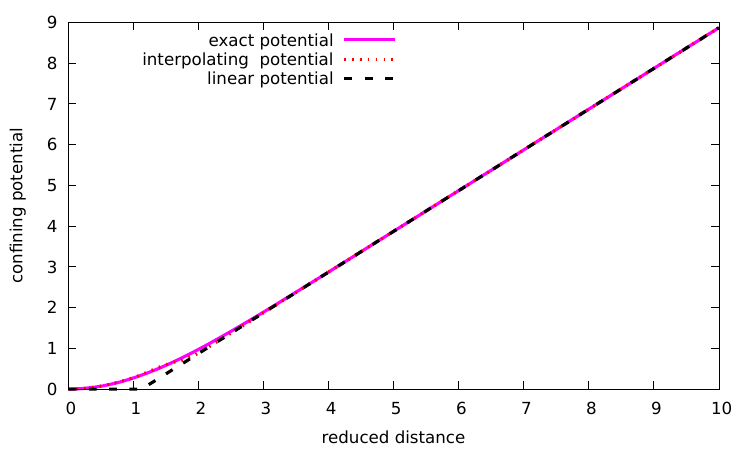}
  \caption{ Solid line: the confining interaction (in units of $\sqrt{\sigma}$), with the constant shift omitted, as a function of the reduced distance, $R/T_g$. The figure also shows the long-range linear part (dashed line) of the confining interaction and the analytic form (dotted line) interpolating between the quadratic short-range and linear long-range pieces. The calculation was performed with $T_g= 0.286 \,\mathrm{fm}$ and $\eta=\sqrt{\sigma}T_g=0.615$. }
  \label{CONFA}
\end{figure}

\begin{figure}[H]

  \includegraphics[width=0.8\textwidth,angle=0]{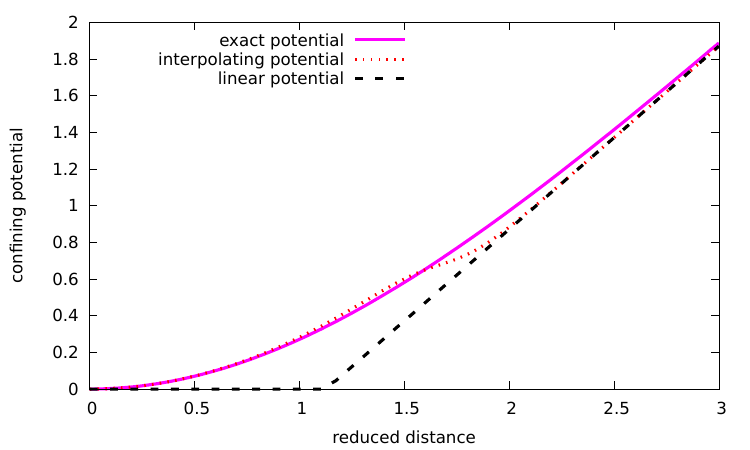}
  \caption{ Zoom-in on the confining potential limited to $R<3/\sqrt{\sigma}\simeq 1.5$\, fm. 
}
  \label{CONFB}
\end{figure}

 This confining interaction is displayed in Figures \ref{CONFA}  and \ref{CONFB} for typical values of the parameters. 
\vspace{-12pt}
\subsubsection{The Bound-State Equation}
In this section, we
give  the bound-state equation for a constituent quark with mass $M=\bar{\mathcal{S}}(\rho)$ moving in a BCS-like state representing the broken (modified) QCD vacuum and submitted to the confining potential with the origin at the string junction point. All the details for the derivation  of the results given below will be given in a long forthcoming paper (referred as [NJLFCM] \cite{NJLFCM}).  
The eigenvalue equation for the confined bound state described by the Dirac wave function, $\left(\psi_n\right)({\bf{x}})$, has the form:
\begin{equation}
\left[-i\bfvec{\alpha}\cdot\vec\nabla_{\bf{x}}+ \beta M\right] \left(\psi_n\right)({\bf{x}}) +  \int d{\bf{z}}\,\beta\,M({\bf x}, {\bf z})\,\left(\psi_n\right)({\bf{z}})  =E_n \left(\psi_n\right)({\bf{x}})
\end{equation}\\
where the mass operator is \cite{KNR2017,NJLFCM}:
\vspace{-9pt}
\begin{adjustwidth}{-\extralength}{0cm}
\centering 
\begin{eqnarray}
M({\bf x}, {\bf y})&=& - C_F\,\,\gamma_0\,\Lambda_{red}({\bf x}, {\bf y})\gamma_0\,\left(V_C({\bf X} +V_C({\bf Y})\right)\nonumber\\
- \Lambda_{red}({\bf x}, {\bf y})&=&\int\frac{d^3 k}{(2\pi)^3}\, e^{-i {\bf k}\cdot ({\bf x} - {\bf y})}\Lambda_{red}({\bf k})=\int\frac{d^3 k}{(2\pi)^3}\, e^{-i {\bf k}\cdot ({\bf x} - {\bf y})}\,\frac{s_k\,- \,c_k\,\bfvec{\gamma}\cdot{\bf{k}}}{2},
\end{eqnarray} 
\end{adjustwidth}
and the chiral angle $\varphi_k$ is defined by $s_k\equiv sin\varphi_k=M/E_k$, $c_k\equiv cos\varphi_k=k/E_k$ and $E_k=\sqrt{k^2+M^2}$. The reduced projector $\Lambda_{red}$ is the difference between  projectors in the positive- and
negative-energy solutions of the Dirac equation. In Ref.  \cite{KNR2017}, a kernel with the same structure as Equation \eqref{STRUCT} was used  (but in the limit of infinitely thin width, $T_g\to 0$). With such a kernel structure,  the solution of this Dirac equation has the form of a quasi-plane wave state:
\begin{eqnarray}
\left(\psi_n\right)({\bf{x}})&=&\int\frac{d^3 p}{(2\pi)^3} \,  e^{-i {\bf p}\cdot {\bf x}}\, (\tilde\psi_n)({\bf{p}})\nonumber\\
 (\tilde\psi_n)({\bf{p}})&=&\sqrt{\frac{1}{2}}\left(
\begin{array}{c}
	\sqrt{1+s_p}\,\,\Phi_n({\bf{p}})\\
	\sqrt{1-s_p}\,\,\bfvec{\sigma}\cdot \hat{\bf p}\,\,\Phi_n({\bf{p}})
\end{array}\right)
=\sqrt{\frac{E_P + M}{2 E_P}}\left(
\begin{array}{c}
	\Phi_n({\bf{p}})\\
	\frac{\bfvec{\sigma}\cdot {\bf p}}{E_P + M}\,\,\Phi_n({\bf{p}})
\end{array}\right).
\end{eqnarray}

We are looking for normalized single-quark states, $|n;ljm\rangle$, with well-defined parity and total angular momentum, and the associated  spinor wave function reads:
\begin{eqnarray}
\Phi_{n;ljm}({\bf{p}})&=&\sqrt{4}\pi\,R_{lj}(p)\,\Phi_{ljm}(\hat{\bf{p}})\nonumber\\
\Phi_{ljm}(\hat{\bf{p}})&=&\sum_{\mu s} Y_{l\mu}(\hat{\bf p})\,\chi_s\,\langle l\mu, \frac{1}{2}s|jm\rangle\quad \hbox{with} \quad \bfvec{\sigma}\cdot\hat{\bf{p}}\,\Phi_{ljm}=-\Phi_{l'jm},\, l'=2j-l.
\end{eqnarray}

The bound-state equation finally reduces to a Schr\"{o}dinger-like equation, 
\begin{eqnarray}
&&E_p\,\Phi_{n;ljm}({\bf{p}})\,+\,\frac{1}{2}\,\int\frac{d^3 q}{(2\pi)^3}\,\tilde{W}_S({\bf p} - {\bf q})\,\bigg(\sqrt{1+s_p}\sqrt{1+s_q}\nonumber\\
&&+\sqrt{1-s_p}\sqrt{1-s_q}\,\bfvec{\sigma}\cdot {\bf p}\,\bfvec{\sigma}\cdot {\bf q}\bigg)\Phi_{n;ljm}({\bf{p}})=\left(E_{jl}\,+\,\frac{2\sigma T_g}{\sqrt{\pi}}\right)\,\Phi_{n;ljm}({\bf{p}}),
\end{eqnarray}
where $\tilde{W}_S({\bf p})$ is the Fourier transform of an effective one-body confining potential $W_S({\bf R})$ such that 
\begin{equation}
W_S\left({\bf R}\right)=\frac{\sigma}{2\,\sqrt{\pi}\,T_g}\,{\bf R}^2
\,I\left({\bf R},{\bf R}\,;\,0\right)=\,\frac{\sigma}{2\,\sqrt{\pi}\,T_g}\,{\bf R}^2
\,\int_0^1 dv\,\int_0^1 dw\,e^{-\left(\left(v-w\right)^2\frac{{\bf R}^2}{4 T^2_g}\right)},
\end{equation}
with asymptotic behavior:
\begin{equation}
R<<T_g :  \quad 
W_S\left({\bf\ R}\right)=\frac{\sigma}{2\,\sqrt{\pi}\,T_g}\, R^2, \qquad
R>>T_g : \quad W_S\left({\bf R}\right)= \,\sigma\,\left(R\,-\,\frac{2}{\sqrt{\pi}}\,T_g\right).
\end{equation}

Actually, this potential deviates from the above $V_C({\bf R})$ by a factor of $1/C_F$ with the constant shift $2\sigma T_g/\sqrt{\pi}$ removed. The bound-state energy can also be written as
\begin{eqnarray}
E_{jl}&=&\int d\bfvec{x}\, \left(\psi_n\right)(\bfvec{x})\left[-i\bfvec{\alpha}\cdot\vec\nabla_{\bfvec{x}}\,+\, \beta M \, +  \,\,W_S(\bfvec{x})\right]\,\left(\psi_n\right)(\bfvec{x})-\,\frac{2\sigma T_g}{\sqrt{\pi}}\nonumber\\
&\equiv & E_{jl, kin}+E_{jl, pot}, \nonumber\\
E_{jl, kin}&=&\int \frac{d\bfvec{p}}{(2\pi)^3}\,E_p\,R^2_{jl}(p), \nonumber\\
E_{jl, pot}&=&\int d\bfvec{r}\, W_S(r)\,\left(|G_{1l}(r)|^2 + |G_{2l'}(r)|^2\right)\,-\,\frac{2\sigma T_g}{\sqrt{\pi}},
\end{eqnarray}
with
\vspace{-9pt}
\begin{adjustwidth}{-\extralength}{0cm}
\centering 
\begin{equation}
  G_{1l}(r)=\int \frac{d\bfvec{q}}{(2\pi)^3}\,\sqrt{1+s_q}\,R_{jl}(q)\,j_l(qr),\quad
G_{2l'}(r)=\int \frac{d\bfvec{q}}{(2\pi)^3}\,\sqrt{1-s_q}\,R_{jl'}(q)\,j_{l'}(qr).  
\end{equation}
\end{adjustwidth}

We look for a Gaussian solution for the lowest orbital with $j=1/2,\, l=0,\,l'=1 $, 
\begin{equation}
R_0(p)= (2\pi)^{\frac{3}{2}}\,\left(\frac{b^2}{\pi}\right)^{\frac{3}{4}}\,e^{-b^2\,p^2/2},  
\end{equation}
with $b$ being a variational parameter. One can search for a solution for any value $M (s)$ of the in-medium constituent quark mass, hence producing a density-dependent nucleon mass as a function of the value of the ``nuclear physics sigma meson'' field $s$ \cite{NJLFCM}. Here, we only need the solution for the vacuum case.
With the QCD inputs given above, $\sigma=0.18\,\mathrm{GeV}^2$ and $\mathcal{G}_2=0.025\,\mathrm{GeV}^4$, yielding   $T_g= 0.286\,\mathrm{fm}$, $\eta=\sqrt{\sigma}T_g=0.615$ and $M_0=356.7$ MeV, the value of $b$  minimizing the orbital energy is $b=0.978/\sqrt{\sigma}$.  The resulting contribution to the quark  orbital energy  is $E_0=E_{0, kin} + E_{0, pot}=1.458\,\sqrt{\sigma}\,+\,0.636\,\sqrt{\sigma}$. The nucleon wave function is just the product of the three quark orbitals properly projected onto the color singlet state with $I=J=1/2$. After center-of-mass correction, the quark core contribution to the nucleon mass is:
\begin{equation}
M_N^{(core)}=\sqrt{9E_0^2 -\langle P^2\rangle} = 3\,\sqrt{E_0^2 -\frac{1}{2 b^2}}=3.62\,\sqrt{\sigma}.
\end{equation}

As a side remark, it is possible to perturbatively incorporate within the model (one obtains $g_A=1.24$)  the pion  cloud and the chromomagnetic contribution to the nucleon mass to obtain a nucleon mass very close to the physical nucleon mass. We will discuss this point in a forthcoming paper \cite{NJLFCM}  (see also \cite{Chanfray2023}), but here we consider only the quark core contribution to extract the response parameters, which can obtained  either semi-analytically or numerically. One finds 
\vspace{-9pt}
\begin{adjustwidth}{-\extralength}{0cm}
\centering 
\begin{equation}
g_S=\frac{M_0}{F_\pi}\left(\frac{\partial M_N^{(core)}(\mathcal{S})}{\partial\mathcal{S} }\right)_{\mathcal{S}=M_0}=6.52,\qquad
C=\frac{M_0^2}{2 M_N}\left(\frac{\partial^2 M_N^{(core)}(\mathcal{S})}{\partial\mathcal{S}^2 }\right)_{\mathcal{S}=M_0}= 0.32.
\end{equation}
\end{adjustwidth}

The value of the $C$ parameter looks a priori small, but since $g_S$ is significantly smaller than $M_N/F_\pi$, its effective value entering the parameters $\Tilde{C}_3$ \eqref{eq:THREEBODN} and $\Tilde{C}_L$ \eqref{eq:LATTIX} is larger, e.g., $(M_N/g_S F_\pi)\, C =0.50$. If we include the value of $C_\chi=0.488$, also derived from the microscopic FCM approach \cite{Chanfray2023}, one has:
\begin{equation}
 \Tilde{C}_3=\frac{M_N}{g_S F_\pi}\,C +\frac{1}{2}\,C_\chi=0.75,\qquad
 \Tilde{C}_L=\frac{M_N}{g_S F_\pi}\,C +\frac{3}{2}\,C_\chi=1.23.
\end{equation}

It turns out that the $s$ field whose value lies in the range $[0 \div-F_\pi]$ does not exactly coincide with the canonical scalar field, $s_c$, of the bosonized NJL Lagrangian. According to~\cite{Chanfray2007}, they are related by 
$s=z_S\,s_c$, where the dimensionless rescaling factor $z_S$ is expressible in terms of the NJL loop  integrals. With the above NJL parameters, its numerical value is $z_S=1.284$. Hence, the values of the scalar coupling constant and of the scalar mass associated with the canonical scalar field are: 
\begin{equation}
 g_\sigma =z_S\,g_S=8.37, \qquad  m_\sigma =z_S\, M_\sigma=919\,MeV.
\end{equation}

This is the reason for which we quote those values of $g_\sigma$ and $m_\sigma$ in the parameters entering the NN potential in Equation \eqref{CHIRPOT}. In addition, notice that the ratio $g_S/M_\sigma=g_\sigma/m_\sigma$ is not affected when passing  from the field $s$ to the canonical scalar field $s_c$, and this rescaling has no effect at the Hartree level but may have a small effect on the Fock terms.
It is also possible to calculate the core rms and the vertex form factors in the various (scalar, axial and vector) Yukawa channels. Here, we give the cutoff values for the equivalent monopole form factors regularizing the corresponding Yukawa coupling to the nucleon:
\begin{equation}
 \Lambda_\sigma=0.86\,GeV,\qquad   \Lambda_\pi=1.01\,GeV,\qquad  \Lambda_v=1.13\,GeV
\end{equation}

This simple estimate just demonstrates that they are close to the $\Lambda =1\,\text{GeV}$ introduced above in Equation \eqref{CHIRPOT}.
\section{Equation of State and Correlations from the QCD-Connected Chiral Confining~Model}\label{sec4}
To prepare the nuclear matter calculations, let us summarize the origin of the various input  parameters distinguishing between those coming from the QCD-connected model and those coming from hadron phenomenology:
\begin{itemize}
    \item Concerning the parameters entering the $NN$ interaction, the scalar and pionic sectors are entirely given or strongly constrained by the QCD-connected model: $g_\sigma=8.37,\, m_\sigma=919\,\text{MeV},\, \Lambda_S=1\,GeV,\,g_A=1.26,\, F_\pi=92.4\,\text{MeV},\, m_\pi=140\,\text{MeV},\, \Lambda_\pi=1\,\text{GeV}$. Notice that at this level, the cutoff parameters $\Lambda_S,\,\Lambda_\pi $, as well as the cutoff in the vector channel $\Lambda_V$, are only approximately  compatible with the QCD-connected model.\\
    The vector-(Lorentz) tensor $NN$ sector is constrained by well-established hadron phenomenology: $g_V=7.5,\, m_V=783\,\text{MeV},\,g_\rho=g_V/3,\, m_\rho=770,\,\text{MeV},\,\kappa_\rho=6$. Notice that in a version of the underlying NJL including pion-axial mixing (i.e., nonvanishing $G_2$), $g_V$ should be promoted to the rank of a QCD-connected parameter. We again stress that a large value of $\kappa_\rho$ (used in the Bonn potential \cite{BONNPOT})  is required to decrease the (Wigner) tensor force without obtaining  too large a D-state probability in the deuteron.
    \item The two parameters entering the three-nucleon force, $C=0.32$ and $C_\chi=0.488$,  are given by the QCD-connected model.   
    \item The parameter $q_C=670\, \text{MeV}$ entering the pair correlation function, $f(r)=1-j_0(q_c r)$, is obtained from the above $NN$ interaction with an adapted G-matrix calculation preserving the UV regularization of the loop integrals entering the correlation energy (see below). Finally, a rather large value of the cutoff, $\Lambda_\rho=2\,\text{GeV}$, for the tensor coupling of the rho meson is needed to obtain a sufficiently large value of the Landau--Migdal parameter $g'$. This is the only constraint or requirement  from nuclear matter phenomenology.
    \item At variance with two parallel works \cite{Rahul,Cham}, the lattice parameters  $a_2, a_4$
    are not taken as inputs but are used to test the consistency of the model.
\end{itemize}

\subsection{Binding Energy of Nuclear Matter}\label{sec4.1}
In a previous paper \cite{Chanfray2007} devoted to a discussion of the  chiral properties of nuclear matter, we calculated the  equation of state with the inclusion of pion (and rho) loops on top of the Hartree mean-field approximation using the first version of the model, i.e., with the L$\sigma$M chiral potential. The energy density is written as:
\begin{equation}
{E\over V}=\varepsilon=\int\,{4\,d^3 p\over (2\pi)^3} \,\Theta(p_F - p)\,E^*_p(\bar s)
\,+\,V_\chi(\bar s)\,+\,{g^2_V\over 2\, m_V^2}\,\rho^2\,+\,\varepsilon^{Loop},
\end{equation}
where $\varepsilon^{Loop}=\varepsilon^{Fock}\,+\,
\varepsilon^{Corr}$ summarizes the loop energy coming from the Fock terms and multi-loop correlations (with pion + rho + short-range  spin--isospin effective interaction). In passing, we recall that the correlation piece (see Equation \eqref{CORREL} below) calculated in the ring approximation actually includes the long-range RPA correlations but with a spin--isospin interaction modified by the short-range correlation.  The calculation presented here is essentially the same as the one in \cite{Chanfray2007} but with the following modifications:
\begin{itemize}
    \item The L$\sigma$M potential (Equation (\ref{eq:VLSM})) is modified by incorporating the factor $1-C_\chi$ into the cubic $s^3$ term with the value $C_\chi=0.488$ obtained in the underlying QCD-connected NJL model, derived from the FCM approach. The parameters entering the in-medium nucleon mass, $g_S$ and $C$, are those given above by the same underlying microscopic model, whereas in \cite{Chanfray2007}, $g_S=M_N/F_\pi$ was taken from the original L$\sigma$M formulated at the nucleonic level, and $C$ was a free parameter.
    \item The parameters governing the various interaction vertices  are those listed in \linebreak Equation~(\ref{CHIRPOT}). In particular, the parameters relative to the scalar sector ($g_\sigma$ and $m_\sigma$) and the cutoffs regularizing the various vertices are given by or compatible with the underlying FCM approach.
    \item The contribution of multi-pion (and multi-rho) exchange is explicitly included in the correlation energy, which thus incorporates  the effect of the two-pion (and two-rho) exchange which is not reducible to iterative processes with NN intermediate states. The effect of these diagrams with at least one $\Delta$ in the intermediate state (see Figure 
 \ref{BUBBLE}) was taken into account in  the construction of the $NN$ potential (Section \ref{sec2.3}, Equation~\eqref{SIGPRIM})  through the introduction of a second scalar field  $\sigma'$ by adding  the Lagrangian $\mathcal{L}_{\sigma'}=-g_{\sigma'}\Bar{N}\sigma' N+\partial^\mu\sigma'\partial_\mu\sigma'/2 -m^2_{\sigma'}\,{\sigma'}/2$ with $g_{\sigma'}=4.8$ and $m_{\sigma'}=550\,MeV$. At the mean-field level, one has $\bar{\sigma}'=-g_{\sigma'}\rho_S/m^2_{\sigma'}$. 
    The net effect of this new scalar field is the modification of the Dirac scalar mass given by Equation (3) in \cite{Chanfray2007}, according to
    $$M^*_N(\Bar{s})=M_N +g_S \Bar{s} +\frac{1}{2}\kappa_{NS}\Bar{s}^3 +g_{\sigma'}\Bar{\sigma}', $$
    where $\bar{s}$ is the solution of the in-medium gap equation. However,  we do not include the associated energy per nucleon, $E_{\sigma'}/A=g_{\sigma'}\Bar{\sigma}' +m^2_{\sigma'}\Bar{\sigma}'^2/2\,\rho \simeq -46\,(\rho/\rho_0)$ MeV, which is contained as a part of the correlation energy. Numerically,  at the saturation density, the chemical potential is $\mu=\partial\varepsilon/\partial\rho =E/A=-13.6$ MeV, which is very different from the Fermi energy when the $\sigma'$ self-energy is ignored, $E^{no\,\sigma'}_{F}=65$ MeV,  in strong violation of the Hugenholtz--Van-Hove  (HVH) theorem.  The $\sigma'$-field energy   is actually  not very far numerically from the dominant linear part of the correlation energy. It follows that the corresponding self-energy contributing to the chemical potential satisfies $\Sigma_{\sigma'}\simeq\partial\varepsilon_{corr}/\partial\rho $. Accordingly, the Fermi energy becomes  $E^{with\,\sigma'}_{F}=-25$ MeV,  in much better agreement with the HVH theorem.  Consequently, the unique effect  of the $\sigma'$  field is to decrease the in-medium Dirac nucleon mass to a value slightly above $750 $ MeV at  normal nuclear matter density, as in \cite{Chanfray2007}.
    (We  recall in passing that this lack of thermodynamic consistency induces only a small non-relativistic correction to the binding energy at  normal density.) 
    \item For the calculation of the pion and rho  Fock terms (Equations (23) and (24) of \cite{Chanfray2007}) and of the correlation energy (Equations (15)--(20) in \cite{Chanfray2007}), we replace the longitudinal and transverse spin--isospin interaction by the one given in Equations \eqref{GEFFSI1}--\eqref{GEFFSI3} based on the Jastrow ansatz (Equation \eqref{JASTROW}) for the correlation function with $q_c=670$ MeV. This procedure has the nice feature of satisfying the Beg--Agassi--Gal theorem, hence providing a natural UV regularization of the loop integrals. In the calculation of the correlation energy, we take the same effective interaction in the    $N\Delta$ and $\Delta\Delta$ channels, with  the appropriate modification of the coupling constant with the $SU(3)$ flavor prescription for the ratio $R_{N\Delta}=g_{\pi N\Delta}/g_{\pi NN}=\sqrt{72/25}$. This generates a unique value for the Landau--Migdal parameters $g'_{NN}=g'_{N\Delta}=g'_{\Delta\Delta}=0.59$. The rather large value of the cutoff, $\Lambda_\rho=2\,\text{GeV}$, for the tensor coupling of the rho meson in a strong rho scenario, $\kappa_\rho=6$, has been chosen to obtain a sufficiently large value of this Landau--Migdal  $g'$ parameter compatible with phenomenology \cite{Ichimura}.  With the notations of  \cite{Chanfray2007}, one has
\vspace{-9pt} 
\begin{adjustwidth}{-\extralength}{0cm}
\centering 
   \begin{eqnarray}
\varepsilon^{Corr}_L&=&{3\over 2}\int {id\omega d{\bf q}\over 
(2\pi)^4}\big[-\ln\big(1-V_{LNN}\Pi^0_N
-V_{L\Delta\Delta}\Pi^0_\Delta-(V^2_{LN\Delta}-V_{LNN}\,V_{L\Delta\Delta})
\Pi^0_N\Pi^0_\Delta\big)\nonumber\\
& &- V_{LNN}\Pi^0_N-V_{L\Delta\Delta}\Pi^0_\Delta\big]\,\nonumber\\
\varepsilon^{Corr}_T&=&3\,\int {id\omega d{\bf q}\over 
(2\pi)^4}\big[-\ln\big(1-V_{TNN}\Pi^0_N
-V_{T\Delta\Delta}\Pi^0_\Delta-(V^2_{TN\Delta}-V_{TNN}\,V_{T\Delta\Delta})
\Pi^0_N\Pi^0_\Delta\big)\nonumber\\
& &- V_{TNN}\Pi^0_N-V_{T\Delta\Delta}\Pi^0_\Delta\big],
\label{CORREL}
\end{eqnarray}
\end{adjustwidth}
where the second line of the two above expressions corresponds to the subtraction of the mean-field Fock terms. 
    \end{itemize}
 
\begin{figure}[H]

  \includegraphics[width=0.8\textwidth,angle=0]{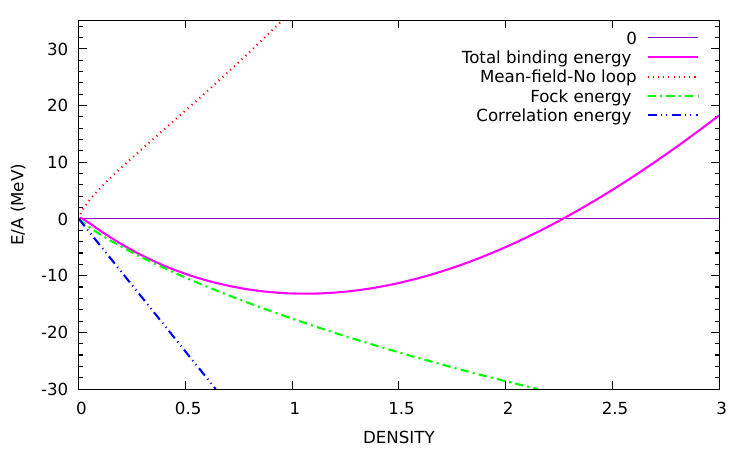}
  \caption{Binding energy 
 of nuclear matter versus $\rho/\rho_0$. Solid line: total energy. Dotted line: mean-field result. Dot-dashed line: spin--isospin Fock term. Double dotted-dashed line: correlation energy}
  \label{EOSFINAL}
\end{figure}

Even if we are aware that each of the various approximations or prescriptions of this multi-step approach should or would deserve a more detailed study, we do not refrain from showing  the results of the calculation without a further fine-tuning of the parameters.  The saturation curve is displayed in Figure \ref{EOSFINAL}. The binding energy at the saturation point is too small, $E_{sat}/A=-13.18$ MeV, and the saturation density is slightly too large, $\rho=1.05\,\rho_0=0.168\,\text{fm}^{-3}$, whereas the incompressibility modulus, $K_{sat}=215$\, MeV, is acceptable, although a bit too low. Even though the saturation point is not perfect, it is rather satisfactory to arrive at decent results for the properties of nuclear matter in an approach based on a microscopic model that mainly  depends on QCD inputs, namely, the string tension and the gluon correlation length (or the gluon condensate) and parameters ($g_V/m_V, \,\kappa_\rho$) known from well-established hadronic phenomenology. A robust conclusion is that the pion and  rho loops  are necessary to obtain sufficient binding, a conclusion already reached in Ref. \cite{Chanfray2007}.  Concerning the comparison with lattice data, our model calculation yields:
\begin{equation}
a_2=1.175\,\text{GeV}^{-1},\qquad   a_4=-0.615\,\text{GeV}^{-3}. 
\end{equation}

The value of $a_2$ is compatible with lattice data and yields a nonpionic contribution to the sigma commutator $\sigma^{(s)}=a_2\,M^2_\pi=23.5$ MeV. 
The value of $a_4$, although a bit high, is essentially compatible with lattice data in the sense that it is much smaller than the value obtained in  the simplest  linear sigma model ignoring the nucleonic response (i.e., $C=0$) and the NJL correction (i.e., $C_\chi=0$), for which $a_4\simeq -3.5\,\text{GeV}^{-3}$. Hence, the model generates the strong compensation required by lattice data from effects governing the three-body repulsive force needed for the saturation mechanism: compare
\mbox{Equations \eqref{eq:LATTIX} and \eqref{eq:THREEBODN}.} 

At variance with our original work \cite{Chanfray2007}, we should also introduce the Fock contribution to the binding energy for the scalar, the  omega and and the (Lorentz vector piece of) rho meson exchanges, i.e., as in Refs. \cite{Massot2008,Cham}.  For orientation, we can  use  a non-relativistic local approximation, replacing the exchange momentum in the meson propagators and vertex form factors by an averaged value  $\bar{q}^2=6 k_F^2/5$, which is valid if the Fermi momentum is small compared to the meson masses:
\begin{eqnarray}
\frac{E_F^{(s)}}{A}&=&\frac{g^2_\sigma}{16}\,\frac{\Lambda_S^2(\bar{q}^2)}{m^2_\sigma+\bar{q}^2}\, \frac{\rho^2\,+\,\rho_S^2}{\rho}\\
\frac{E_F^{(\omega)}}{A}&=&-\frac{g^2_V}{8}\,\frac{\Lambda_V^2(\bar{q}^2)}{m^2_V+\bar{q}^2}\, \frac{2\,\rho_S^2\,-\,\rho^2}{\rho},\quad
\frac{E_F^{(\rho)}}{A}=-\frac{g^2_V}{24}\,\frac{\Lambda_V^2(\bar{q}^2)}{m^2_\rho+\bar{q}^2}\, \frac{2\,\rho_S^2\,-\,\rho^2}{\rho}.
\end{eqnarray}

\begin{figure}[H]

  \includegraphics[width=0.8\textwidth,angle=0]{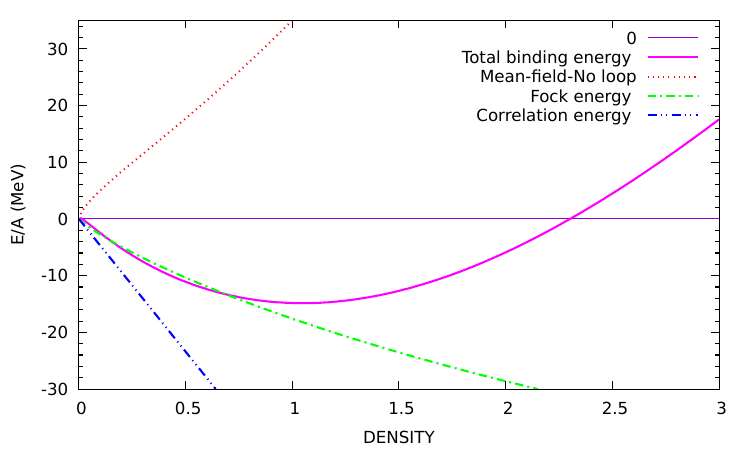}
  \caption{Binding energy 
 of nuclear matter versus $\rho/\rho_0$ with  modified $C=0.4$ and $C_\chi=0.46$ parameters and including scalar and (Lorentz time component of) vector exchange Fock terms. Solid line: total energy. Dotted line: mean-field result. Dot-dashed line: spin--isospin Fock term. Double dotted-dashed line: correlation energy.}
  \label{EOSF}
\end{figure}

In addition, this Fock term is included perturbatively in the sense that we remain in the Hartree basis, since we neglect the Fock contribution to the scalar self-energy of the nucleon. It turns out that there is a strong compensation between the scalar and omega Fock energy, but the rho Fock contribution provides a sizable attraction, and the net result at  normal nuclear density is $E_F/A=-3.31$ MeV. We also introduce the effect of short-range correlations by subtracting from the Hartree and Fock contributions the same quantities but  adding $q^2_C$ to the exchange momentum squared in the meson propagators and form factors. The net result at  normal nuclear matter density is extremely small,  $E_{SRC}/A=-0.03$~MeV. Of course, the effect of these Fock terms  in the presence of SRCs is the significant modification of the saturation properties: the saturation density is now much too large, $\rho_{sat}=1.2\,\rho_0$, whereas the binding energy becomes $E_{sat}/A=-16.9$ MeV.  To solve this problem, one certainly has to first  incorporate the effect of the Fock terms in the nucleon self-energy, hence working in the fully self-consistent Hartree--Fock basis with a resulting modification of the Dirac effective mass. If we decide not to change the parameters that impact the bare NN interaction, one can also advocate 
a possible improvement in the FCM approach, which is flawed with various shortcomings, thus allowing variations in the $C$ and $C_\chi$ parameters. Just to give a flavor of what might result, we modify the parameters governing the repulsive three-body force by taking a value of $C=0.4$, compatible with other confining models,  and  by decreasing the NJL parameter $C_\chi$ by $5\%$.  The resulting saturation point (see Figure \ref{EOSF}) is slightly improved, $E_{sat}/A=-14.85$ MeV, $\rho_{sat}=1.05\,\rho_0=0.168\,\text{fm}^{-3}$, $K_{sat}=230$ MeV, whereas the lattice parameter  $a_4=-0.41\,\text{GeV}^{-3} $ is compatible with the lattice results.

It nevertheless remains true that further  works are needed to adjust the parameters, with the  QCD-connected model acting as a guideline,  to address the question of the EOS at a higher density, particularly for the neutron star EOS. 

  \subsection{The Interrelated Role of the Short-Range Correlations}\label{sec4.2}
  Short-range correlations (SRCs) play a crucial role in the loop energy. The Fock contribution can be split into pion and rho contributions. At normal nuclear matter density, the presence of the SRC reduces  the repulsive pion Fock term from $E_F^\pi/A=11.71$ MeV to $E_F^{\pi + SRC}/A=2.87$ MeV and transforms the repulsive rho Fock term, $E_F^\rho/A=8.86$ MeV, into a strongly attractive contribution, $E_F^{\rho + SRC}/A=-20.5$ MeV. Consequently, the total Fock term, $E_F/A=-17.6$ MeV, appears as an important contribution to the binding energy of nuclear matter. The same conclusion has been reached in a parallel work, where all the Fock terms have been incorporated into a full HF approach \cite{Cham}. The combined effects of the composite nature of the nucleon embedded in the form factors and of the SRC control the magnitude of the RPA correlation integrals. In particular, the UV behavior of the spin--isospin interaction linked to the Jastrow correlation function provides a natural regularization of the loop integral as a consequence of the Beg--Agassi--Gal theorem. It  is interesting to make a comparison with the chiral perturbation
calculation in Ref. \cite{KFW02}, where the effect of a short-range correlation is absent. 
In this  work, the contribution of the iterated pion exchange is found to be $-68$ MeV, and the short-range physics is described by a unique cutoff regularizing the pion loop. This has to be compared with our result where $E_{corr}/A=-32.3$ MeV at  normal nuclear matter density.
This calculation of the correlation energy fully incorporates the recoil correction and the energy dependence of the spin--isospin interaction, which means that the $q^2$ appearing in the pion and rho  propagators  is systematically replaced by $q^2 -\omega^2$, and the integral $\int id\omega d{\bf q}$ becomes $-\int dz d{\bf q}$ after a Wick rotation (see \cite{Chanfray2007,Massot2009}), hence transforming $q^2 -\omega^2$ into $q^2 +z^2$.

\begin{figure}[H]

  \includegraphics[width=0.3\textwidth,angle=270]{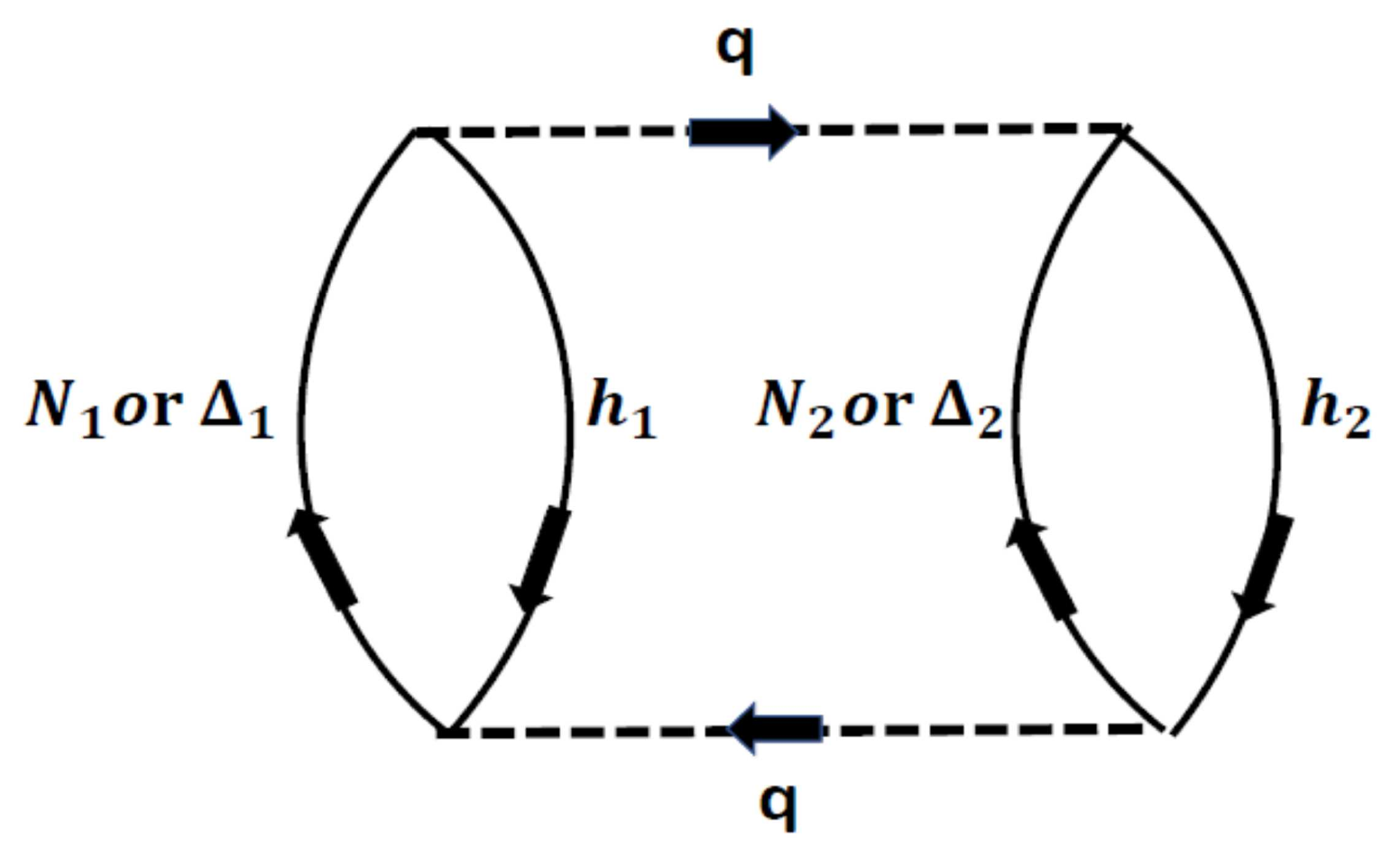}
  \caption{ Diagrams contributing to the  correlation energy or to the depopulation of the Fermi sea.}
  \label{BUBBLE}
\end{figure}

In the following discussion, we consider the case  of $\rho=0.8 \,\rho_0$, i.e., $k_F=245$ MeV, appropriate for the carbon nucleus. The full correlation energy per nucleon is $E_{corr, full}=-27$ MeV.  If we limit the calculation of the correlation energy to the two-loop order, which is essentially equivalent to second-order perturbation theory,
\begin{eqnarray}
\varepsilon^{Corr}_L
&\simeq&{3\over 2}\int {id\omega d{\bf q}\over 
(2\pi)^4}\big[V^2_{LNN}\Pi^{02}_N\,+\,2\,V^2_{LN\Delta}\Pi^0_N\Pi^0_\Delta\,+\,V^2_{L\Delta\Delta}\Pi^{02}_\Delta\big]\nonumber\\
\varepsilon^{Corr}_T
&\simeq&3\,\int {id\omega d{\bf q}\over 
(2\pi)^4}\big[V^2_{TNN}\Pi^{02}_N\,+\,2\,V^2_{TN\Delta}\Pi^0_N\Pi^0_\Delta\,+\,V^2_{T\Delta\Delta}\Pi^{02}_\Delta\big]
,\label{2P-2h}
\end{eqnarray}
its value is slightly increased to $E_{corr}=-30.7$ MeV. At this level, one can clearly distinguish the various contributions 
 to the correlation energy per nucleon, corresponding to the presence of  $NN,\,N\Delta,\,\Delta\Delta$ in the intermediate states (see Figure \ref{BUBBLE}):
\begin{equation}
 E_{corr}=  E_{corr}^{NN}\,+\,E_{corr}^{N\Delta}\,+\,E_{corr}^{\Delta\Delta}= -11.8\,-\,13\,-5.9 =-30.7\,\text{MeV}.
\end{equation}

If we remove the energy dependence of the spin--isospin interaction, i.e., in a static approximation, one can show that this two-loop correlation energy takes the simple form
\begin{equation}
		E_{corr,st}=\frac{1}{2A}\,\sum_{h_1,h_2, p_1, p_2}\frac{\left|<p_1\, p_2\,|\,G_{\sigma\tau}\,|\,h_1\, h_2>\right|^2}{\epsilon_{h_1}+\epsilon_{h_2}-\epsilon_{p_1}-\epsilon_{p_2}},\label{ESTATIC}
\end{equation}    
where $G_{\sigma\tau}$ is the static spin--isospin interaction given by Equations \eqref{GEFFSI1}--\eqref{GEFFSI3}. Apart for the energy denominator, this expression has the same structure as the expression for the mean depopulation of the nucleon Fermi sea discussed above in Section \ref{sec2.2}:
\begin{equation}
		\Delta n=\frac{1}{2A}\,\sum_{h_1,h_2, p_1, p_2}\left|\frac{<p_1\, p_2\,|\,G_{\sigma\tau}\,|\,h_1\, h_2>}{\epsilon_{h_1}+\epsilon_{h_2}-\epsilon_{p_1}-\epsilon_{p_2}}\right|^2 \equiv \kappa.\label{NSTATIC}
\end{equation}    
but omitting the Pauli exchange term, which is consistent with the ring summation. This allows the simultaneous discussion, at the same level of approximation, of the role of the SRC in the loop energy affecting the equation of state and in the depopulation of the Fermi sea, $\Delta n$, which is an excellent indicator of the effect of the SRC on the nuclear response functions and which also coincides with the wound integral of the Brueckner theory.  In the two above expressions, the particle $p$ states can be either a nucleon above the Fermi sea or a $\Delta$ state. The numerical calculation of the RPA-2p-2h correlation energy can be performed by using Equations (48)--(50) in our previous work \cite{CEM2022} by just replacing the squared energy denominators by the (always negative) single-energy denominators. In this static approximation, the various contributions to the long-range correlation energy are: 
\begin{equation}
 E_{corr, st}=  E_{corr, st}^{NN}\,+\,E_{corr, st}^{N\Delta}\,+\,E_{corr, st}^{\Delta\Delta}= -\,11.1\,-\,14.8\,-8.2 =-34.2\,\text{MeV}
\end{equation}
which is slightly above the nonstatic calculation.  As a byproduct, we can also  estimate the
contribution to the correlation energy of the scalar ($s,\,\sigma'$) and vector meson ($\omega$)
exchanges that we have not considered explicitly before. This amounts  to replacing, in the dotted lines of Figure \ref{BUBBLE}, the spin--isospin interaction by the $\sigma+\omega$ interaction in the non-relativistic case; one finds numerically:
\begin{equation}
   E_{corr, st}^{\sigma+\omega}= -3.6 \,\text{MeV},
\end{equation}
which is probably overestimated since we did not incorporate the in-medium dropping of the scalar coupling constant.

We now come to the calculation of the Fermi sea depopulation and the kinetic energy per nucleon as  a complement of the model results given in Subsection 5.2 of \cite{CEM2022}. From baryon number conservation, the total depopulation per nucleon, $\Delta n$, is equal to the fraction of states (nucleon or $\Delta$) outside of the Fermi sea. One obtains:  
	\begin{eqnarray}
		\Delta n &=&\Delta n^N \,+\,\Delta n^\Delta= 0.113\,+\,0.019= 0.132. 
\end{eqnarray}

The proportion of $\Delta$ in the ground state induced by the short-range correlation is thus $2\,\%$. The contribution of the probably overestimated scalar+vector exchange is  smaller than the pion+rho exchange:
\begin{equation}
   \Delta n^{\sigma+\omega}= 0. 059,
\end{equation}
and the full depopulation of the Fermi sea (or the BHF wound integral) is $\Delta n=0.19$,  which is not far from the estimate, $\Delta n \simeq 0.15\div0.2$, of Ref. \cite{Cioffi}. 

The calculation of the (non-relativistic) kinetic energy per nucleon, another test of the role of the  SRC,
can be performed using Equations (51)--(54) in Ref. \cite{CEM2022}. One finds: 
\vspace{-9pt}
\begin{adjustwidth}{-\extralength}{0cm}
\centering 
\begin{equation}
		<t>=\frac{3}{5}\frac{k^2_F}{2M}\,-\,<t>^{depop}\,+\,<t>_{tail}^{N}\,+\,<t>^{\Delta} =19\,-\,2.72\,+\,10.3\,+\,5.9\,= 32.5\,\text{MeV}.
	\end{equation}
\end{adjustwidth}
	
 One can notice the sizable contribution of the deltas present in the ground state. The  contribution of the scalar+vector exchange is small:
 \begin{equation}
		<t>^{\sigma+\omega}=-1.25\,+\,3.95=2.7\,\text{MeV}.
	\end{equation}
	
 The total kinetic energy is $<t>= 35.2$ MeV, where a significant part  $16\, \text{MeV}$ comes from SRCs depopulating the Fermi sea. This value is also within  the estimate, 
 $<t>\simeq35~\text{MeV}$, of  \cite{Cioffi,Cioffi1,Cioffi2}, which is needed to explain the EMC effect.
 
 \section{Conclusions and Perspectives}\label{sec5}
 The chiral confining model, with inputs  linked to genuine QCD quantities (string tension, gluon condensate) and well-established hadronic phenomenology ($\kappa_\rho\sim 6$), is able to generate, via an adapted G-matrix approach, a pair correlation function that ensures the natural UV regularization of the pionic  correlation energy. It also gives, without further fine-tuning, reasonable results for the chiral properties of the nucleon ($a_2$ and $a_4$ parameters extracted from lattice data) and saturation properties of nuclear matter. In addition, it allows the simultaneous discussion  of the role of SRCs in the loop energy affecting the equation of state and in the depopulation of the Fermi sea, $\Delta n$, which is an excellent indicator of the effect of the SRC on the nuclear response. It is remarkable  that one obtains  good results for $\Delta n$ and the kinetic energy per nucleon $<t>\sim 35$ MeV, which is needed to reproduce the celebrated EMC effect. Conversely, one can remark that these values of the Fermi sea depopulation and of the kinetic energy per nucleon induced by the SRC are compatible with a large value of the rho meson tensor coupling ($\kappa_\rho\sim6$ or $c_\rho\sim 2$) and of the associated cutoff ($\Lambda_\rho\sim 2\,GeV$), which, in turn, induces a large binding energy coming from pion and rho loops. This very interesting link can be captured by just looking at Equations \eqref{ESTATIC} and \eqref{NSTATIC}.
 
 Concerning the QCD-connected chiral confining model specifically, which is at the heart of the whole approach, the effect of the NJL-like enriched chiral effective potential significantly improves the compatibility of the lattice data with the model calculations of the dimensionless response parameter $C$.  Another important result discussed in detail in~\cite{Chanfray2023} is the existence of a particular combination of $C$ and $C_\chi$ ($\Tilde{C}_L$) constrained by LQCD, and  a closely related combination ($\Tilde{C}_3=C+C_\chi/2$) governing the repulsive  three-nucleon force ensuring the saturation mechanism. Hence, the fundamental QCD theory and nuclear matter modeling are linked by, on the one hand, the LQCD data $a_2$ and $a_4$ and, on the other hand, what we have called the ``QCD connected parameters'', namely, the response parameters $g_s$ and $C$. Indeed, these results provide a link between the chiral properties of the nucleon and the saturation mechanism and/or a link between the fundamental theory of strong interaction and nuclear matter properties, although the results presented in this paper  should be enriched in various aspects of this multi-step approach.
 
 The above discussion opens perspectives that may impact the strategy to be adopted in the near future. First, as pointed out in Section \ref{sec4.1}, the Fock terms, including the rearrangement terms \cite{Massot2008}, have not been consistently incorporated  at the level of nucleon self-energy.  Although this problem is of minor importance for the binding energy around the nuclear matter density, the use of a Hartree--Fock basis in place of the Hartree basis for the nucleon Dirac wave function may play a very important role at a high density in limiting the maximum mass of a hyperonic neutron star, as pointed out in Ref.~\cite{Massot2012}. The passage to this Hartree--Fock basis can be implemented without formal difficulty  but at the expense of increasing  numerical complexity \cite{Massot2008,Cham}. However, the incorporation of this thermodynamic consistency (HVH theorem) for the correlation energy certainly necessitates  additional formal developments.  Second,  there are still  improvements to be made in several directions of the theoretical treatment of the FCM-QCD model, as discussed and suggested in Section \ref{sec3.3}. 
 Given the remaining theoretical uncertainties, one possible strategy is to use a Bayesian method with QCD-connected parameters and lattice data ($a_2, a_4$) parameters, implemented with their associated uncertainties,  as   prior input variables. This may be implemented with the methodology used in two recent papers of the Lyon group \cite{Cham, Rahul}.

\vspace{6pt} 



 \authorcontributions{ 
 Conceptualization, G.C., M.E. and M.M. ;
 methodology, G.C., M.E. and M.M.; 
 software, G.C..; 
 validation, G.C., M.E. and M.M.; 
 formal analysis, G.C..; 
 writing---original draft preparation, G.C.; 
 writing---review and editing, G.C., M.E. and M.M.  
 All authors have read and agreed to the published version of the manuscript.
}



\acknowledgments{One of us (G.C.)  acknowledges his collaborators in the Lyon group, H. Hansen, J. Margueron and M. Chamsedinne, for fruitful discussions and collaborations for part of this work.}

\conflictsofinterest{ 
The authors declare no conflict of interest.
}

\begin{adjustwidth}{-\extralength}{0cm}

\printendnotes[custom]
\reftitle{References}

\PublishersNote{}
\end{adjustwidth}
\end{document}